\documentstyle[12pt,aaspp4]{article}

\def\etal{{\it et~al.\/\ }}

\def\teff{{\it T$_{\rm eff}$}}
\def\logg{{\rm log~$g$}}

\def\kms{{km~s$^{-1}$}}  
\def\dexkpc{{dex~kpc$^{-1}$}}  
\def\hg{{H$\gamma$}}

\def\Rm31{{R$_{\rm M31}$}}
\def\fei{{\ion{Fe}{1}}}
\def\feii{{\ion{Fe}{2}}}
\def\mgi{{\ion{Mg}{1}}}
\def\mgii{{\ion{Mg}{2}}}
\def\magdex{{mag\,dex$^{-1}$}}
\def\R23{{R$_{23}$}}

\slugcomment{In preparation for the Astrophysical Journal}

\lefthead{Venn et al.}
\righthead{M31 A-F Supergiants Abundances}

\begin{document}

\title{Analysis of Four A-F Supergiants in M31 \\
       from Keck HIRES Spectroscopy 
	\footnote{Based on observations obtained at
	the W.\,M.\,Keck Observatory, which is operated
	as a scientific partnership among the California
	Institute of Technology, the University of
	California, and the National Aeronautics and 
	Space Administration.  The Observatory was made
	possible by the generous financial support of the
	W.\,M.\,Keck Foundation.} }

\author{K. A. Venn\altaffilmark{2,3} }
\affil{Macalester College, Saint Paul, MN, 55105}

\author{J. K. McCarthy\altaffilmark{4} }
\affil{Palomar Observatory, California Institute of Technology, 
Pasadena, CA, 91125}

\author{D. J. Lennon\altaffilmark{2} }
\affil{Isaac Newton Group of Telescopes (ING), 
Santa Cruz de La Palma, Canary Islands, E-38780, Spain}

\author{N. Przybilla and R. P. Kudritzki\altaffilmark{5}}
\affil{Institut f\"ur Astronomie und Astrophysik, Universit\"at Sternwarte
       M\"unchen, M\"unchen, D-81679, Germany}

\and

\author{M. Lemke\altaffilmark{6}}
\affil{Dr.\,Karl Remeis-Sternwarte, Bamberg, D-96049, Germany}

\altaffiltext{2} {Institut f\"ur Astronomie und Astrophysik, 
Universit\"at Sternwarte M\"unchen, Scheinerstrasse 1, M\"unchen, 
D-81679, Germany}
\altaffiltext{3}{Adjunct Assistant Professor, Department of Astronomy,
University of Minnesota, Minneapolis, MN, 55455 } 
\altaffiltext{4}{Present address: Pixel Vision, Inc., Advanced
Imaging Sensors Division, 4952 Warner Avenue, Suite 300, Huntington Beach,
CA, 92649 }
\altaffiltext{5}{Max Planck Institut f\"ur Astrophysik, 
Garching, D-85740, Germany} 
\altaffiltext{6}{Present address: INA-Werk Schaeffler, Herzogenaurach,
Germany} 

\begin{abstract}

The first stellar abundances in M31 are presented, based on 
Keck I HIRES spectroscopy and model atmospheres analyses of 
three A-F supergiants, 41-2368, 41-3712, and A-207.  
We also present the preliminary analysis of a fourth star,
41-3654.  We find that the stellar oxygen abundances are in good agreement 
with those determined from nebular studies, even though the
stars do {\it not} show a clear radial gradient in oxygen. 
The uncertainties in the stellar abundances are smaller than the range 
in the nebular results, making these stars ideal objects for
further studies of the distribution of oxygen in M31.
We show that the stars can be used to study the abundance distributions 
of other elements as well, including iron-group and heavier elements.

The A-F supergiants also provide direct information on
the metallicity and reddening of nearby Cepheid stars.  
We have examined the metallicity and reddening assumptions used
for Cepheids within 1' of our targets and noted the differences 
from values used in the literature.

\end{abstract}


\section{Introduction}

Light element abundances in the disk of M31 are known 
primarily from nebular analyses of \ion{H}{2} regions and 
supernovae remnants (Blair \etal 1981, 1982, 
Dennefeld \& Kunth 1981), and more recently from a 
few disk planetary nebulae (2 PNe by Jacoby \& Ciardullo 1999, 
3 by Jacoby \& Ford 1986, and 2 by Richer \etal 1999).
The nebular abundances have identified a mild gradient 
in oxygen with galactocentric distance, with a fairly
large range in the abundances at a given radial distance.
Until recently, the 
stars in M31 have been too faint to observe with high
resolution spectroscopy for detailed atmospheric analyses.
The advent of large telescopes and efficient detectors is
changing this though, and now high resolution spectroscopy 
of stars fainter than V=17 can be obtained, making stellar 
atmospheric  analyses of stars beyond the Magellanic Clouds 
a new possibility.  

One question that new elemental abundances from stars can 
address concerns abundance gradients that are commonly identified 
in spiral galaxies (surveys of \ion{H}{2} regions in numerous 
spiral galaxies have been carried out by McCall \etal 1985, 
Vila-Costas \& Edmunds 1992, and Zaritsky \etal 1994).   
While there is no doubt about the existence of these gradients
in many galaxies, 
it is the exact form of the gradient, the dispersion in abundance 
at a given galactocentric radial distance, and the constancy of 
the gradient between different elements, that new stellar studies 
can address.   And, in contrast to abundance determinations using 
\ion{H}{2} regions, which provide information mainly on light elements 
(e.g,. He, N and O), stellar analyses can provide the abundances of 
these and heavier elements (e.g., O, Mg, Fe, and possibly even Ba). 
Through the study of these additional elements, new 
insights will be gained into the nucleosynthetic history 
and chemical evolution of spiral galaxies.

Stellar atmosphere analyses are also relevant to Cepheid
distance determinations, through metallicity calculations 
and reddening estimates.  As the accuracy in the determination 
of H$_o$ reaches the stated HST Key Project goal of 10\% 
(Kennicutt, Freedman, \& Mould 1995), then   
the systematic uncertainties in the Cepheid distances themselves
become increasingly important.  In particular, the question
of metallicity effects on the Cepheid period-luminosity (PL)
relationship, and uncertainties
in the reddening estimates (c.f., Kennicutt \etal 1998). 
The A-F supergiants are physically linked to Cepheids; 
these stars have similar young ages and intermediate-masses, 
and should have very similar compositions and galactic (thin disk) 
locations.
The element ratios (e.g, O/Fe) found from the A-F supergiants 
can be used for the most accurate metallicity/opacity 
studies of Cepheids, e.g., simple scaling of the oxygen abundances 
from nebular analyses to all other elements does not allow 
for differences due to the chemical evolution history of different 
galaxies.
The similar locations of A-F supergiants and Cepheids also mean that 
reddening estimates can be checked or improved (i.e., after 
detailed atmospheric analyses of the A-F supergiants).

In this paper, we present the first elemental abundances
from high resolution stellar spectroscopy of individual stars 
in M31.  Model atmospheres analyses for three A-F supergiants 
are presented, as well as preliminary results for a fourth star.
The elements observed include alpha-elements (O, Mg, Si, Ca, Ti)
as well as iron-group elements (Cr, Fe, Ni), and the first heavy 
element abundances (Y, Ba, Ce, Nd) for M31.  The oxygen abundances
are compared to the nebular radial gradient, and we discuss the
importance of examining other elemental gradients as well as
element ratios to study the chemical evolution of the disk of M31. 
We also determine reddening to the A-F supergiants, and compare
the adopted reddening and metallicity estimates of nearby Cepheids 
to discuss potential improvements in the Cepheid distance determinations.

\section{Observations and Reductions}

Observations of three A-type supergiants (41-3712, 41-2368, and 41-3654) 
in M31 were taken on August 1, 1995, with the $10\,$m Keck I telescope and 
HIRES spectrograph (Vogt \etal 1994). 
Two 45-minute exposures of each star were made in sub-arcsecond seeing
conditions through a 1.1-arcsec slit, giving $R=35,000$ over a 4 pixel
resolution element.
A combined signal-to-noise ratio $\ge40$ per pixel, or 
$S/N>80$ per resolution element, was attained after coaddition.
Similar observations of the F-supergiant (A-207) were taken on August 4, 1995.
Weather conditions only allowed for two exposures, one for 50 minutes and a
second for 30 minutes;
since this star is fainter than the A-supergiants, the overall
signal-to-noise is somewhat lower, $S/N\sim50$ per resolution element.
The wavelength range spanned $4300\,$\AA\ $\leq \lambda \leq 6700\,$\AA\ in
30 echelle orders, although the wavelength coverage was not complete for
$\lambda >5000\,$\AA\ on the TK2048 CCD\ used. The FWHM of the Keck images,
as measured in the CCD spectra perpendicular to the dispersion, 
was $0.85\,$arcseconds (dominated by atmospheric seeing), or
$2.2\,$pixels after on-chip binning by $2\times$ in the spatial
direction. Slit length was limited to $7.0\,$ arcseconds on the sky to
prevent overlapping orders at the short wavelength extreme.
Such high quality spectra had only been possible for Galactic
and Magellanic Cloud stars before.

The two-dimensional CCD echelle spectrograms were reduced at Caltech using
a set of routines written for echelle data reduction (c.f., 
McCarthy \etal 1995, McCarthy \& Nemec 1997) 
within the FIGARO package. Even though the stars are isolated 
and not within \ion{H}{2} regions, the stellar HIRES spectra revealed
significant broad (over $300$\,\kms) nebular contamination
of the Balmer line profiles, easily recognizable to the sides of the 
stellar spectra in the two-dimensional CCD data.
This nebular contamination was removed from the stellar spectra prior to
extraction by fitting low-order polynomials to ``sky apertures'' adjacent to
the stellar spectrum in the spatial direction on the slit (also discussed 
in McCarthy \etal 1995 for analysis of individual stars in M33).

\begin{figure}
\plotone{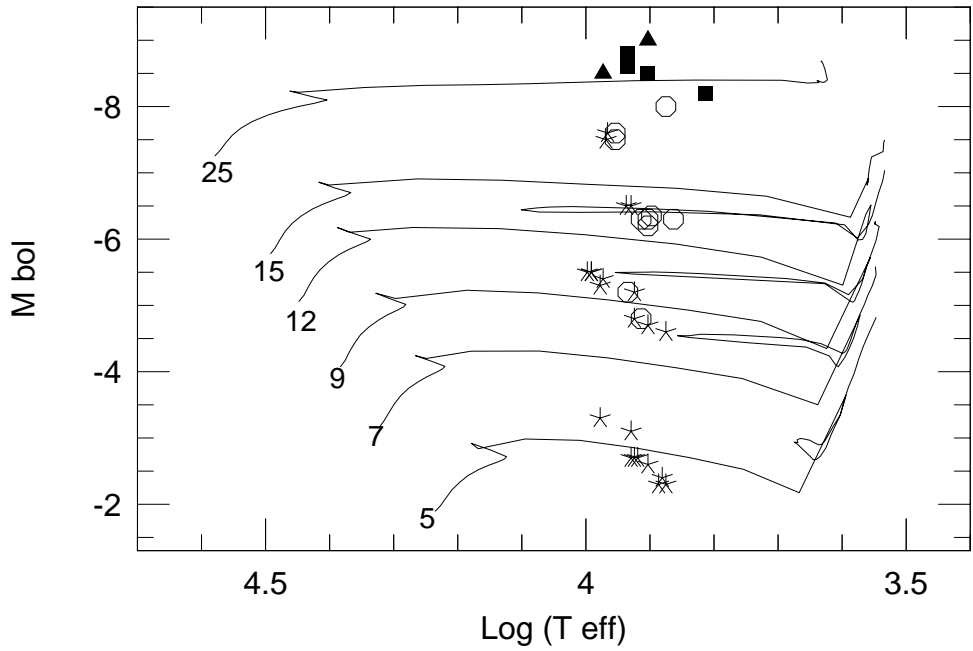}
\caption{An HR diagram to show the approximate location 
of the four M31 A-F supergiants in this paper ({\it filled squares}).   
Also shown are 22 A-supergiants in the Galaxy ({\it asterisks}), 
10 in the SMC ({\it empty circles}), and 2 in M33 ({\it filled triangles})
for which detailed atmospheric analyses have been published
(Venn 1995b, 1999, McCarthy \etal 1995).   
Stellar evolution tracks from Schaller \etal (1992) are shown, 
with stellar masses labelled. 
\label{fig-hrd}}
\end{figure}

\begin{figure}
\plotone{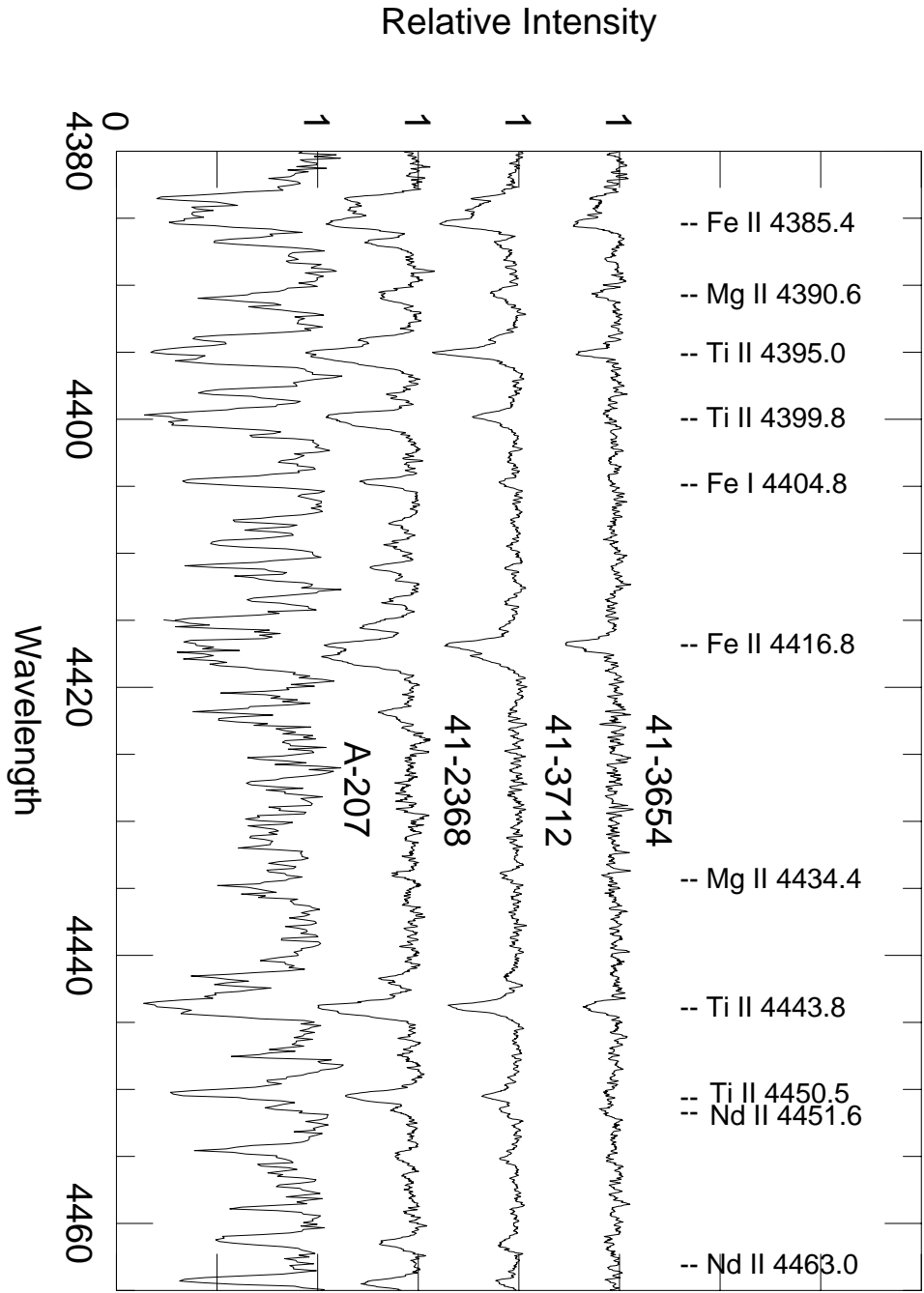}
\caption{Sample spectra near 4400~\AA.  A few lines are
identified, particularly \mgii\ lines used in the atmospheric
analyses, as well as two \ion{Nd}{2} features in A-207.  
\label{spec4400}}
\end{figure}

\begin{figure}
\plotone{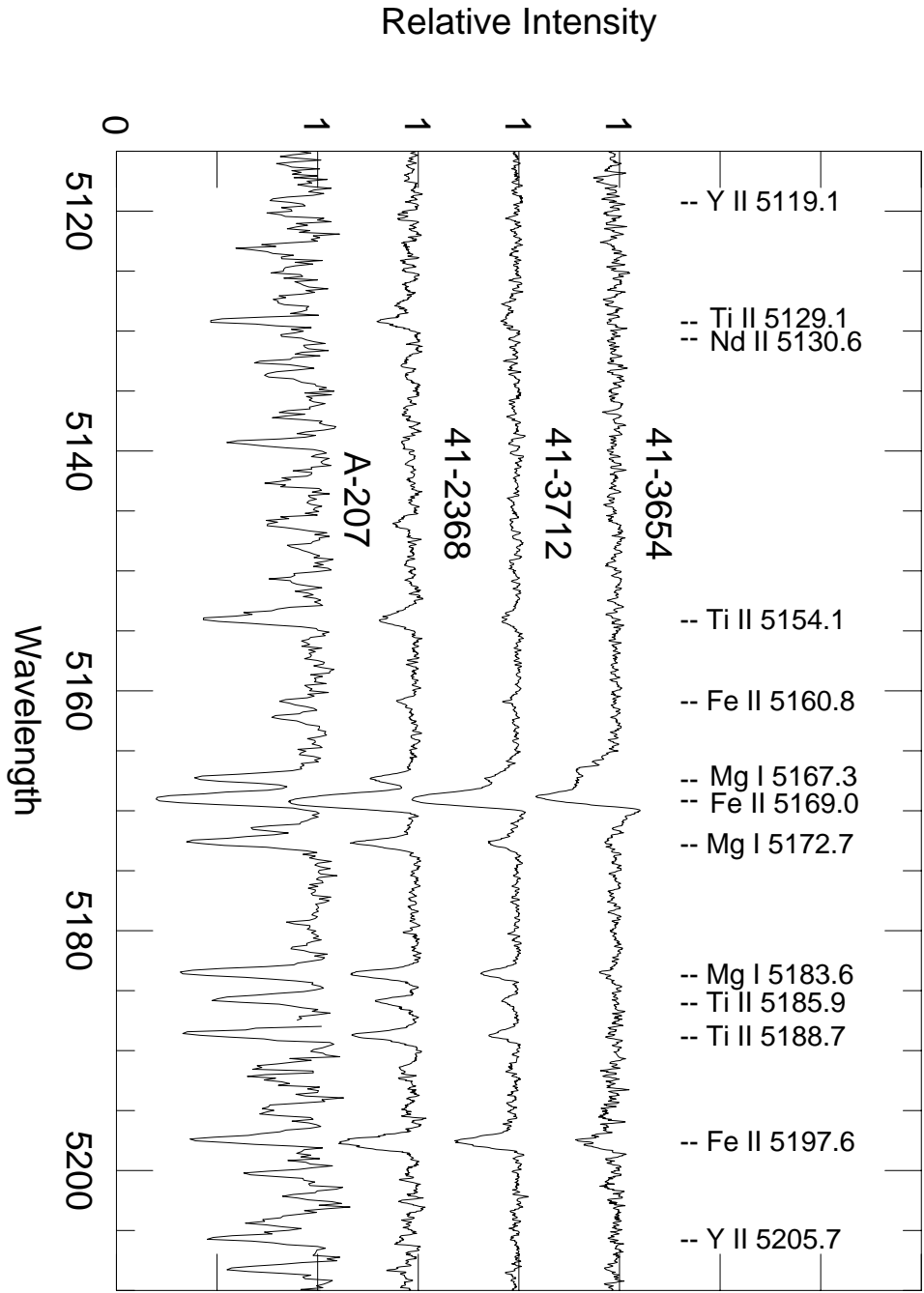}
\caption{Sample spectra near 5200~\AA.  A few lines are
identified, particularly \mgi\ lines used in the atmospheric
analyses, as well as a \ion{Nd}{2} and two \ion{Y}{2} features
in the cooler stars.
\label{spec5200}}
\end{figure}

Three targets (41-3712, 41-2368, \& 41-3654) 
were selected from the low resolution survey 
of luminous blue stars in M31 by Humphreys, Massey \& Freedman (1990).
Additionally, 41-3712 and 41-3654 were confirmed as isolated, normal
A-type supergiants from low resolution spectroscopy by Herrero \etal (1994). 
An additional target (A-207) was selected from Humphreys (1979) because
of its location in Baade's Field IV.  [Note, that A-207 is in the 
western-most of three small associations identified in Plate IV
of Baade \& Swope (1963), and it is not the star labelled ``207'' in 
Figure~2 of Humphreys (1979), which is instead B-207].
Coordinates and $UBV$ colors are listed in Table~\ref{atms}.   
Their estimated locations on an HRD are shown in Fig.~\ref{fig-hrd}.
Sample spectra are shown in Figs.~\ref{spec4400} and \ref{spec5200}.

\section{Atmospheric Analyses} \label{atms-analysis}

The M31 A-F supergiant photospheres have been analysed using
ATLAS9 (hydrostatic, line-blanketed, plane parallel) model 
atmospheres (Kurucz 1979, 1988).
These atmospheres have been used successfully for photospheric
analyses of A-F supergiants in the Galaxy and Magellanic Clouds
(Venn 1995a, 1995b, 1999, Luck \etal 1998, Hill 1997, Hill \etal 1995).   
However, as seen in Fig.~\ref{fig-hrd}, 
the stars analysed here are more luminous than
previously studied A-F supergiants (by Venn 1995a, 1999), 
which poses a few new challenges.  

Firstly, these stars have stronger radiation fields such that departures
from LTE can be expected to increase.   An examination of LTE-grey,
LTE-line-blanketed, NLTE-grey, and NLTE-partially-blanketed
models has been carried out by Przybilla (1997).  He has 
shown that these atmospheres are remarkably similar deep in the
photosphere where the continuum forms.   In fact, the largest
effect in this atmospheric region is caused by neglecting line-blanketing.
Thus, we have elected to use the fully line-blanketed ATLAS9 
model atmospheres as the best representation of the photospheric 
continuum and deep line forming regions.  

Secondly, the stronger radiation fields can create velocity fields
and a stellar wind.   41-3654 and 41-3712 have strong stellar winds, 
as seen by their H$\alpha$ P~Cygni profiles in Fig.~9 and 12 in
McCarthy \etal (1997).  The stellar wind in 41-3654 is much stronger 
than that in 41-3712, seen empirically by comparing the heights of the 
H$\alpha$ P~Cygni emission peaks, and the fact that H$\beta$ still has 
a P~Cygni profile in 41-3654, but not 41-3712. 
P~Cygni profiles can also be seen in some very strong 
Fe~II lines, e.g., see \ion{Fe}{2} 5169 in Fig~\ref{spec5200}. 
Only weak lines that form deep in the photosphere are used in
the supergiant analyses, which are not usually sensitive to velocity 
fields.   In the case of 41-3654, we know the velocity field 
affects the photosphere and can affect our analysis
(discussed below), but the photosphere of 41-3712 is not affected 
by its wind (see McCarthy \etal 1997). 

Thirdly, one may question whether these stars are spherically
extended.  Calculations of the atmospheric thicknesses 
(between $\tau_{\rm 5000}$=2/3 and 0.001)
and the stellar radii (based on M$_{\rm bol}$ and 
effective temperature,
see Venn 1995b) show that atmospheric extension is $\sim$3~\% 
for three stars, 41-2368, 41-3712, and A-207.  Extension
for the fourth star, 41-3654, may be as large as 10\%,
which is a significant amount and is discussed further below.

A summary of the atmospheric parameters determined here for the
stars in M31 are listed in Table~\ref{atms}.   The methods
for these determinations are discussed individually below. 

\subsection{M31-41-3712 \& M31-41-2368 Atmospheres} \label{astars} 

\begin{figure}
\plotone{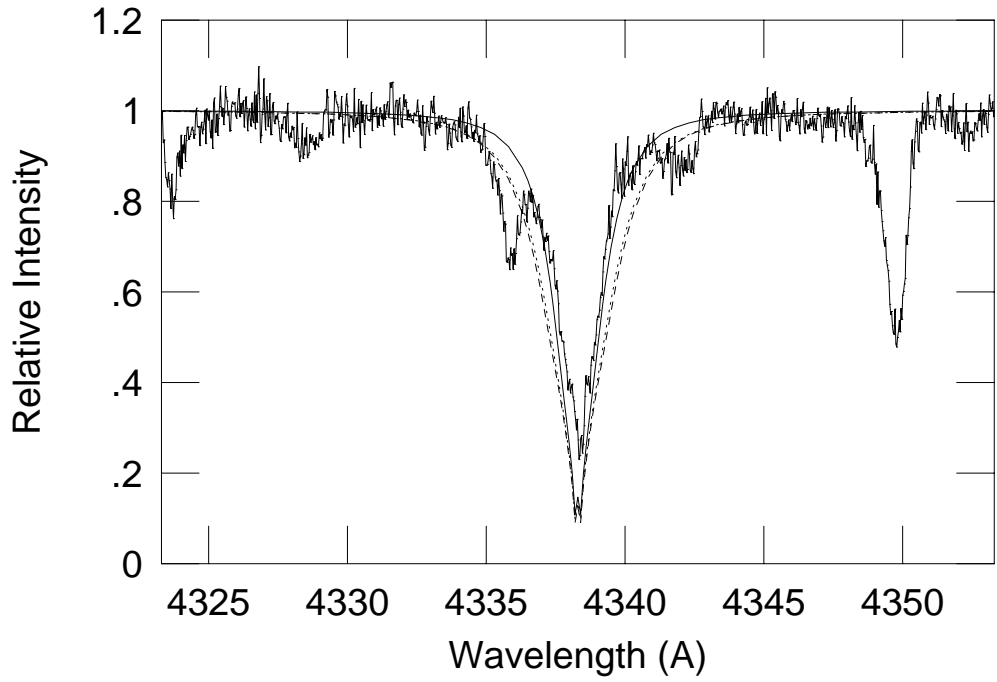}
\caption{H$\gamma$ line fits for 41-3712, including the best fit 
from a model with \teff=8400, \logg=0.9 ({\it solid line}), as well
as \teff=8200, \logg=0.9 ({\it dashed line}) and \teff=8400, \logg=1.0
({\it dotted line}).  
\label{m3712-hg}}
\end{figure}

\begin{figure}
\plotone{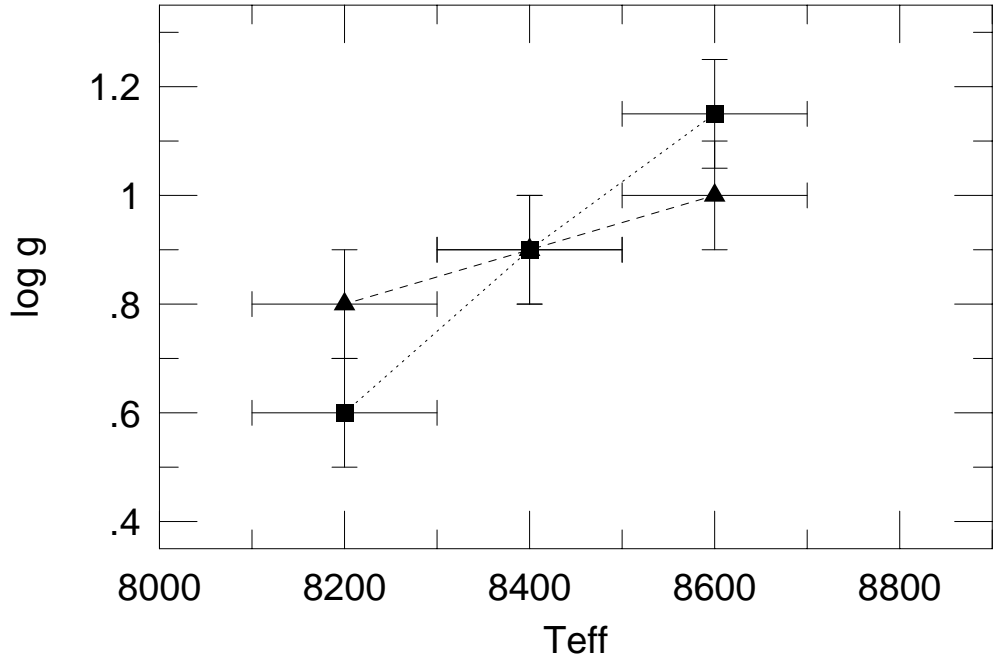}
\caption{Atmospheric parameter determination for 41-3712.
\mgi/\mgii\ is shown by {\it filled squares} connected by a {\it dotted line}.
\hg\ fit parameters are shown by {\it filled triangles} connected by
a {\it dashed line}.  The model parameters at the intersection point have
been adopted for this analysis. 
\label{m3712-atm}}
\end{figure}

Spectral features have been used to determine the model atmosphere
parameters for 41-3712 and 41-2368, in particular the wings of
the \hg\ line profile (e.g., Fig.~\ref{m3712-hg}) 
and ionization equilibrium of \mgi/\mgii\ (e.g., Fig.~\ref{m3712-atm}).  
These features were rigorously examined for normal Galactic 
A-supergiants by Venn (1995b), and can be expected to yield 
reliable parameters to within $\Delta$\teff=$\pm$200\,K, 
$\Delta$\logg=$\pm$0.1.  Both 41-3712 and 41-2368 appear to
be normal A-type supergiants, thus this analysis method is
appropriate for these two stars. 
 
Examination of H$\alpha$ shows that 41-2368 has a very weak
wind affecting only the core of the line, whereas 41-3712 
exhibits a significant P~Cygni profile indicating a substantial wind.
McCarthy \etal (1997) have analysed the wind of 41-3712 in detail, 
and found a mass loss rate of \.M=1.1 $\pm$0.2 x10$^{-6}$ M$_\odot$/yr. 
However, their analysis has also shown that the deeper layers
of the atmosphere are unaffected by the wind, which has only a 
mild influence on H$\beta$ and is almost negligible for H$\gamma$. 
Nevertheless, the blue wing of \hg\ was primarily scrutinized for 
atmospheric parameter constraints, see Fig.~\ref{m3712-hg}. 

NLTE corrections have been calculated for 
\mgi\ and \mgii\ in both stars, and included in the ionization
equilibrium calculations.   The corrections
are negligible for the \mgii\ lines in both stars 
($\Delta$log(Mg\,II/H)=$\le$0.05), 
while the corrections for the log(\mgi/H) abundances range from 
+0.02 to +0.23~dex, correlated with line strength. 
Lines stronger than 200\,m\AA\ were neglected since
these lines form over several atmospheric layers, including
some at small optical depths.
The Mg NLTE calculations use the Gigas (1988) \mgi/\mgii\
model atom and a system of programs first developed by
W.~Steenbock at Kiel University and further developed and
upgraded by M.\,Lemke.   Mg NLTE calculations in 
Galactic A-F supergiants have been described by Venn (1995b). 

There are two ionization states of iron available in these stars
as well, however equilibrium of \fei/\feii\ is not used for
atmospheric parameter determinations in these stars.  These
stars are moderately-warm A-type supergiants where NLTE effects
on \fei\ lines are expected to be large, 0.2 to 0.3~dex 
(c.f., Boyarchuk \etal 1985, Gigas 1986).   \fei\ is overionized 
by the UV radiation field in these stars, but this has a
negligible effect on \feii\, which is the dominant ionization
state.  Thus, the same authors predict that NLTE effects are 
negligible for \feii\ lines, also confirmed by more recent 
detailed calculations by Becker (1998).

Microturbulence ($\xi$) was found by examining the line 
abundances for \ion{Ti}{2}, \ion{Cr}{2}, \feii\ 
(and \fei\ for 41-2368), and requiring no relationship 
with equivalent width.  Allowing for an uncertainty in 
$\Delta\xi$ of $\pm$1\,\kms\ brings the results from the
different species into very good agreement.

\subsection{M31-A-207 Atmosphere} \label{A207}

A-207 is a much cooler star than 41-2368 and 41-3712,
with no weak \mgi\ lines useful for atmospheric parameter
determinations.  Thus, the atmospheric parameters for 
A-207 were determined by comparing the results from 
other spectroscopic, and photometric, indicators.

{\it $UBVR$ colors:}
Firstly, the $UBVR$ colors for A-207 were compared to theoretical
calibrations by Bessell \etal (1998), after an {\it a~priori} 
estimate of reddening from its spectral type.  
$UBV$ colors are reported in Table~\ref{atms}.
Color indices are sensitive to temperature in F-type stars, 
however our results for ($B-V$), ($U-B$), and ($V-R$) are 
not in good agreement with each other.
For example, from the Bessell \etal (1998) calibrations, 
($B-V$) implies \teff=6250 at \logg=0.5, whereas ($U-B$) 
suggests \teff=8250 at \logg=1.0 (these are the lowest
gravity models reported).  Also, ($V-R$) is degenerate 
with \teff\ ranging from 5000 to 9000\,K near \logg=1.0.  
The range in these color calibrations is as large as the 
A-F temperature range.  Thus, the atmospheric parameters 
for A-207 were selected from spectral features.  

\begin{figure}
\plotone{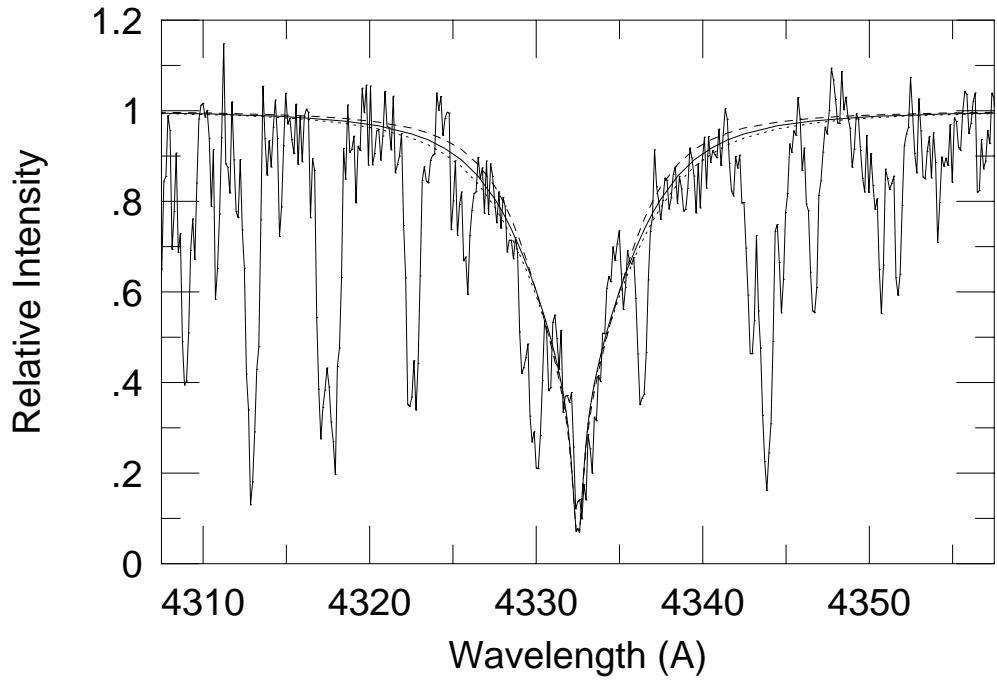}
\caption{H$\gamma$ line fits for A-207, including the best fit from
a model with \teff=6700, \logg=0.2 ({\it solid line}), as well as
\teff=6700, \logg=0.3 ({\it dotted lines}), and \teff=7000, \logg=0.3
({\it dashed line}).
\label{a207-hg}}
\end{figure}

{\it \hg\ wings:}
The \hg\ profile was examined (as was done for the other 
stars in this paper), and yet most fits are rather poor for 
A-207.  This is partially due to the quality of the spectrum
in the bluest orders, 
as well as the significant metal-line blending.  
The best fits to \hg\ in A-207 are shown in Fig.~\ref{a207-hg};
fits are strongly \teff sensitive, but only weakly gravity sensitive.
 
{\it \fei\ versus $\chi$:}
Another spectroscopic parameter is the \fei\ line abundance 
versus lower excitation potential ($\chi$).
This has been used successfully as a temperature indicator
for F-supergiants in the past (e.g., Luck \etal 1998, 
Luck \& Lambert 1992).    This proved to be the
least reliable indicator for our dataset though, primarily
because of the high scatter in the line abundances.
Even after the spectrum was careful scrutinized to search
for potential blends and after the atomic data was carefully
reviewed, the full range in the best 41 \ion{Fe}{1} lines 
(out of over 200 measured features) was 
$\Delta$log(\fei/H) $\sim$1.0~dex.
Negligible trends in log(Fe/H) vs $\chi$ were seen 
in the range of 6500 to 8500~K (for low gravity atmospheres),
the same range found from the $UBV$ colors.

{\it \fei\ versus \feii:}
The average \fei\ and \feii\ abundances were examined for
ionization equilibrium.  For F-supergiants, temperature is usually
selected from one of the methods above, and iron ionization
equilibrium is used only for gravity (c.f., Luck \etal 1998, 
Hill 1997, Luck \& Lambert 1992).   Since no method 
above produced a well defined temperature, then \fei/\feii\
is used here to examine both temperature and gravity (as \hg\ 
was used).   Throughout the elemental abundance analyses
(discussed further below), strong lines are eliminated from
the line list;  for A-207, a limit of W$_\lambda\le$160~m\AA\
was adopted.  Nevertheless, the scatter in the \fei\ and \feii\ 
abundances is large, $1\,\sigma\,\sim$0.3 and 0.2~dex, respectively.
This scatter is similar to that found in other F-G supergiant 
analyses (e.g., Hill 1997, Luck \etal 1998, Luck \& Lambert 1992,
Russell \& Bessell 1989). 
Finally, a locus of \teff-gravity pairs that reproduce Fe 
ionization equilibrium was calculated for A-207.
 
\begin{figure}
\plotone{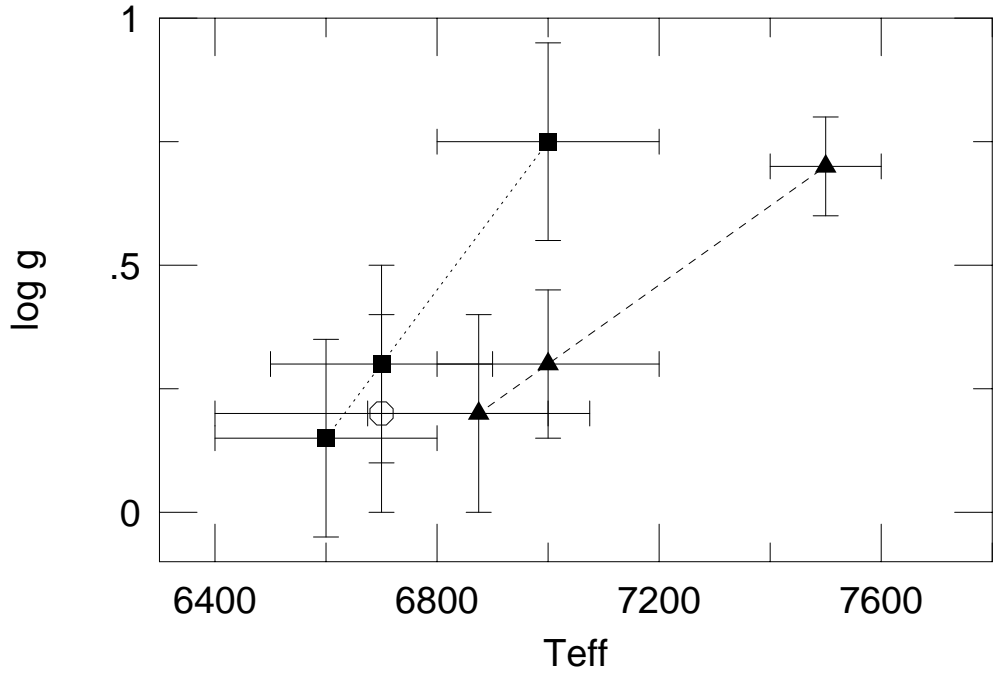}
\caption{Atmospheric parameter determination for A-207. 
\mgi/\mgii\ is shown by {\it filled squares} connected by a {\it dotted line}.
\hg fit parameters are shown by {\it filled triangles} connected by
a {\it dashed line}.  Model parameters adopted for this analysis are
shown by the {\it empty circle} with the estimated uncertainties.
\label{a207-atm}}
\end{figure}

{Parameters for A-207:}
The final atmospheric parameters for A-207 have been selected
as \teff=6700 $\pm$300~K and \logg=0.2 $\pm$0.2, 
with $\xi$=8 $\pm$2~\kms.   
The loci of iron ionization equilibrium and H$\gamma$ \teff-gravity
pairs are shown if Fig.~\ref{a207-atm}.
The uncertainty in $\xi$ in this star is larger
than for the others in this paper because of the large scatter 
in the line abundances.

\subsection{M31-41-3654 Atmosphere} \label{m3654}

Standard spectroscopic features for the analysis 
of an A-type supergiant have been observed in 
41-3654 (e.g., \hg, \mgi, and \mgii\ features
are observed in the spectrum, see Figs.~\ref{spec4400}
and \ref{spec5200}),
however we consider our analysis of this 
star only preliminary in this paper. 
This star has a very strong stellar wind, examined in
detail by McCarthy \etal (1997), which is expected
to impact the line forming region of the photosphere,
unlike the A-F supergiants previously analysed (e.g.,
Venn 1995b, Luck \etal 1998). 
This will affect the pressure and temperature stratification
in a model atmosphere, and it means that most of the 
standard assumptions in the ATLAS9 models will not apply, 
e.g., hydrostatic equilibrium and LTE.

In this preliminary analysis, we simply examine the
\hg\ profile, and \mgi/\mgii\ equilibrium.
No standard ATLAS9 model could be generated to fit the 
H$\gamma$ line profile, i.e., the lowest gravities were 
still too large.  The observed H$\gamma$ line profile
forms over several layers in a stellar atmosphere, although
the wings form primarily in the photosphere which is why
it is usually a useful photospheric diagnostic.  A stellar
wind can contribute to H$\gamma$ (line filling due to
wind emission), but also incoherent electron scattering 
can also fill the line wings (see discussion by 
McCarthy \etal 1997).  Thus, H$\gamma$ has only limited
use for gravity determinations in the most luminous
supergiants. 
More specifically, ATLAS9 fails to fit the 41-3654 
H$\gamma$ line profile because the radiative 
pressure due to the line opacity is too strong 
in this star (i.e.,  g$_{rad}$~$>$~g), thus hydrostatic 
equilibrium breaks down and there is an outward acceleration, 
i.e., a stellar wind.  McCarthy \etal (1997) 
used unified model atmospheres 
(calculated according to Santolaya-Rey \etal 1997) that 
include spherical geometry, radiative equilibrium, and NLTE
radiative transfer in the comoving frame, but not line
blanketing, and they managed to fit the observed H$\gamma$ 
profile with \teff=8900 and \logg=0.9. 
The unified models are ideal for a stellar winds analysis, 
but not for an analysis of the stellar photosphere,
e.g., since they neglect line blanketing.
 
Further tests with ATLAS9 showed that Mg ionization
equilibrium could be attained with parameters
ranging from \teff/gravity = 8500/0.8 to 9100/1.3 
(using the NLTE corrected abundances).
All of these models result in near solar 
Mg but significantly depleted iron-group elements,
log(Fe,Cr,Ti/H)$\sim-0.5$.
The Mg and iron-group results are not significant 
though since an appropriate ATLAS9 model atmosphere
could not be generated.   The stellar wind effects
(and possibly NLTE) effects on the atmospheric 
structure (not accounted for in ATLAS9) will affect
the photospheric abundance determinations. 
Firstly, if the model atmosphere structure is
distorted by NLTE and stellar wind effects, then
this would affect the Mg ionization equilibrium
since the \mgi\ and \mgii\ features form at slightly 
different optical depths.   For example, 
\ion{Mg}{1} $\lambda$5183 has $\chi$=2.72~eV
and forms near log($\tau_{\rm 5000}$)$\sim-1$,
while \ion{Mg}{2} $\lambda$4390 has 
$\chi$=10.00~eV and forms near 
log($\tau_{\rm 5000})\sim-0.4$.
Similarly, the \ion{Ti}{2} lines form in
a similar location to \mgi\, whereas the
\ion{Cr}{2} and \ion{Fe}{2} lines tend to
form in deeper layers. 
Secondly, a simple calculation of the atmospheric extension 
for 41-3654 is $\sim$5-10~\% (between $\tau_{\rm 5000}$=2/3 
and 0.001) based on the range of atmospheric parameters
estimated for this star.  A study of ATLAS models by 
Fieldus \etal (1990) found that 10~\% extension in 
A-supergiants could weaken moderately-strong 
($\sim$100\,m\AA) \ion{Fe}{2} lines by $\sim$10\%.
An additional important effect is the influence of
incoherent electron scattering resulting from the 
large extension of the photosphere (discussed by McCarthy \etal 1997)
which would weaken the lines even further.
We have tried scaling our equivalent widths up by
10~\% and find that the abundances increase,
$\Delta$log($X$/H)= +0.1 to 0.2.

Finally, we note that the helium abundance in an A-supergiant 
atmosphere can also have an effect on its structure 
(c.f., Kudritzki 1973, Humphreys, Kudritzki \& Groth 1991).  
Test calculations show that changing helium from 
9~\% (ATLAS9 standard) to 20~\% in an A-supergiant atmosphere
increases the Balmer jump and Balmer line strengths.
This is because an increase in helium affects the 
mean molecular weight of a column of gas, which 
affects the opacity and pressure stratification. 
But also, helium is mostly neutral in A-type stars,
thus reducing the number of free electrons available 
for Thompson scattering (the dominant opacity source 
at these temperatures). 
Unfortunately, \ion{He}{1} lines are only observed in 
early A-type stars, and NLTE analysis of these lines
have provided inconsistent helium abundances in the
past (c.f., Husfeld 1994).   Thus, determination of
helium abundances in specific stars is uncertain.  
However, our test calculations (increasing helium from
9 to 20~\%) also show that the effects are essentially 
identical to an increase in gravity 
($\Delta$\logg$\sim$+0.2) for 41-3712, e.g., the 
Balmer line profiles 
and the metal line abundances are {\it identical}. 
On the other hand, this is not the case for 41-3654; 
a change in both gravity and temperature are indicated 
to maintain Mg ionization equilibrium, also the derived 
Mg abundance itself increases.  This is due to larger 
NLTE corrections for both species in this slightly 
hotter star. 

Thus, we consider our analysis of the atmospheric parameters
of 41-3654 as preliminary.  Significant improvement to this
preliminary analysis requires model developments that are 
beyond the scope of this paper.
We include our results in Table~\ref{atms}, but we do not 
consider this star throughout the discussion.

\section{Elemental Abundances} \label{elem-abu}

Elemental abundances have been calculated using both
spectrum syntheses and individual line width analyses. 
All calculations have been done using a modified and
updated version of LINFOR\footnote{LINFOR was original 
developed by H.\,Holweger, W.\,Steffen, and W.\,Steenbock 
at Kiel University.  Since, it has been upgraded and 
maintained by M.\,Lemke, with additional modifications by N.\,Przybilla.}.

Equivalent widths are listed in Table~\ref{lines1}
for the non-iron group elements (as an example).
For all lines, atomic data was adopted from the 
literature; an attempt has been made to adopt, 
(1) laboratory measurements data, e.g., O'Brien 
\etal 1991 for \fei, and
(2) opacity project data, e.g., 
Biemont \etal (1991) for \ion{O}{1}. 
(3) or critically examined atomic data, e.g., NIST
data (c.f., {\it http://physics.nist.gov}).
Semi-empirical values calculated by Kurucz (1988,
also see {\it http://cfa-www.harvard.edu/amdata})
were adopted when necessary. 
Solar abundances are adopted from Grevesse \& Sauval (1998).

Not all spectral lines observed were 
used in this analysis. In particular, strong lines were 
neglected.  In most cases, strong lines include those with 
W$_\lambda$\,$\ge$200\,m\AA, e.g., the line strength where
the uncertainty in microturbulence ($\sim\pm$1~\kms) yields
an change in log($X$/H) $\sim\pm$0.1~dex.   
In some cases, a smaller W$_\lambda$ limit was 
used if large deviations in the abundances were apparent 
after a preliminary analysis; e.g., in A-207, 
only \fei\ lines weaker than 160\,m\AA\ were included.   
Weak lines help to ward off uncertainties in the model 
atmospheres analysis due to NLTE and spherical extension
in the atmospheric structure, as well as NLTE and 
microturbulence effects in the line formation calculations.  

\begin{figure}
\plotone{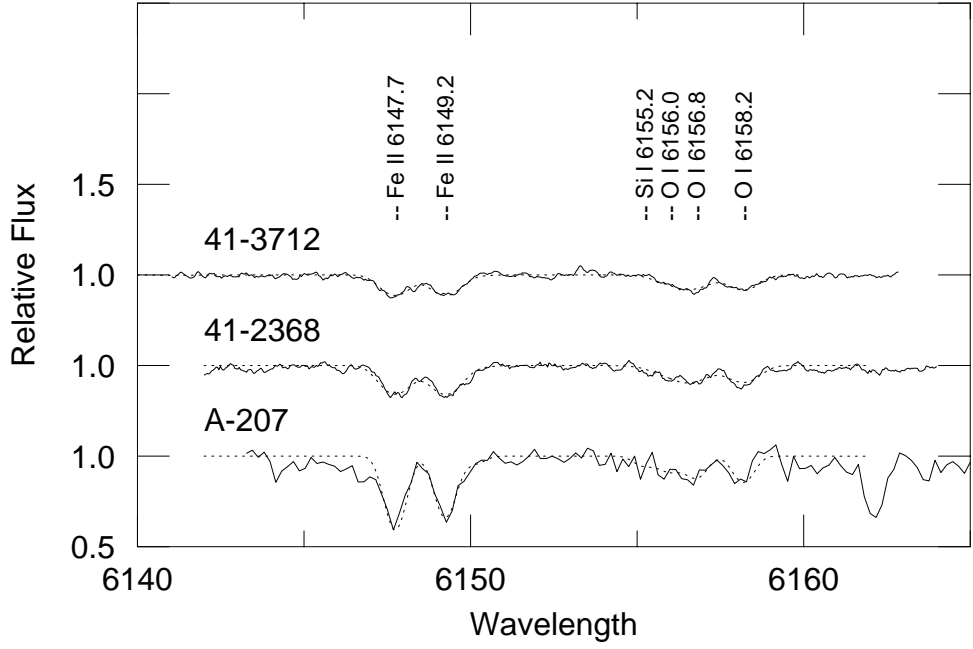}
\caption{Observed \ion{O}{1} 6158~\AA\ feature and spectrum syntheses
({\it dotted line}) for our fully analysed A-F supergiants.   
\label{ospec}}
\end{figure}

Average elemental abundances for each star are listed 
in Table~\ref{abu}.   One sigma errors are
tabulated based only on the line-to-line scatter.   
All oxygen abundances are from spectrum synthesis of the 6158\,\AA\
feature; the observed spectra and synthesis fits are shown
in Fig~\ref{ospec}.  In A-207, the feature is blended with 
\ion{Si}{1}, which was included in the synthesis and its abundance 
was allowed to vary; the oxygen results were nearly insensitive 
to the silicon abundances, e.g., $\Delta$\,log(Si/H) =$\pm$0.3 
caused $\Delta$\,log(O/H) =$\pm$0.02. 
NLTE effects on the \ion{O}{1}6158 feature
are predicted to be small in A-supergiants, and negligible for 
F-supergiants. For example, a correction of 
$\Delta$log(O/H)=$-$0.25 is predicted from detailed NLTE analyses 
(Przybilla \etal 2000, also in good agreement are the earlier
detailed studies by Takeda 1992 and Baschek \etal 1977) 
for the atmospheric parameters of 41-3712.  The predictions 
reduce to $\Delta$log(O/H)=$-$0.20 and 0.0 for 41-2368 and A207, 
respectively.   LTE and NLTE \ion{O}{1} abundances are
listed in Table~\ref{abu}.
 
For all three stars, synthesis of three \feii\ lines 
was done at the same time as the oxygen spectral syntheses 
(i.e., absorption lines at $\lambda$6147, $\lambda$6149, and $\lambda$6150).
The syntheses of these \feii\ lines was used to estimate rotational 
velocities (see Table~\ref{atms}), and the iron abundance results 
were included in the average \feii\ abundances in Table~\ref{abu}. 

Several lines of s- and r-process elements have been observed
(see examples in Figs.~\ref{spec4400} and \ref{spec5200}) and
analysed in two stars.  Note that only one to four lines per 
species have been analysed (see Tables~\ref{lines1} and \ref{abu}), 
making these results somewhat uncertain, e.g., unknown blends, 
quality of the atomic data, unrecognized NLTE effects.   
Several additional lines were observed in A-207, however
their line strengths are well over 200 m\AA\ and are not
considered reliable abundance indicators.   This is especially
true considering their low excitation potentials, indicating
line formation at optical depths above the photosphere.
With respect to NLTE effects on the s- and r-process 
line formation, Lyubimkov \& Boyarchuk (1982) found that Ba 
abundances measured from resonance lines are 0.2 to 0.4~dex 
less than those from subordinate lines in Canopus (an F-type supergiant).   
In this analysis, barium is measured from two resonance lines 
and one subordinate line, however the abundances are in very good
agreement, $\Delta$log(Ba/H)=0.05~dex only.   Thus, either the NLTE
corrections are smaller than predicted for the resonance lines, 
or both resonance and subordinate lines are affected uniformly.
NLTE effects on other s- and r-process elements are not 
currently available.   

To ascertain the reliability of these heavy element abundances,
we have examined the published abundances in Galactic F-supergiants
by Luck \etal (1998).  Luck \etal analysed up to 11 species of
s- and r-process elements in 11 F-G supergiants, including all of the
species in this paper.   They measured one to 18 lines 
per element.  The typical {\it range} in the
\ion{Y}{2}, \ion{Zr}{2}, \ion{Ce}{2}, and \ion{Nd}{2} abundances 
in any one star examined by Luck \etal\ is $\pm$0.1\,dex.   
This strongly suggests that the s- and r-process abundances
can be reliably determined in cool supergiants.   
In this paper, we find the weighted average s- and r-process 
abundances to be [s+r/H]=+0.18 $\pm$0.15 and +0.22 $\pm$0.02 
for A-207 and 41-2368, respectively. 

Abundance uncertainties due to $\Delta$\teff=+200\,K, 
$\Delta$\logg=$-$0.1 are shown for the two A-stars 
in Table~\ref{abu-unc}. Uncertainties due to microturbulence 
($\Delta\xi$=$\pm$1~\kms) are not listed since they are very 
small ($\sim$0.02 to 0.05~dex) in this weak line analysis.  
For A-207, abundance uncertainties due $\Delta$\teff=+300\,K, 
$\Delta$\logg=$-$0.2 and $\Delta\xi=-$2\,\kms\ are shown
in Table~\ref{abu-unc}.

\section{Discussion}

Stellar abundance calculations in M31 can be used to
address several questions:  

(1) Do the stellar oxygen abundances show the same abundance 
gradient in M31 as the nebular oxygen abundances?

(2) What are the abundances of other elements in M31? 
Do these elements also show radial gradients?  What can
differences in the element ratio gradients tell us about
the chemical evolution of M31?

(3) How do the analyses of A-F supergiants in M31 improve the
Cepheid distance determinations through more accurate metallicity 
and reddening determinations?

Each of these will be addressed separately in the 
following subsections.  We stress that three stars are
insufficient to answer any of these questions at this
time; in this discussion, we shall simply show how stellar
abundances in M31 can be used, 
particularly if we can increase the sample size.

\subsection{The Oxygen Abundance Gradient in M31}

\begin{figure}
\plotone{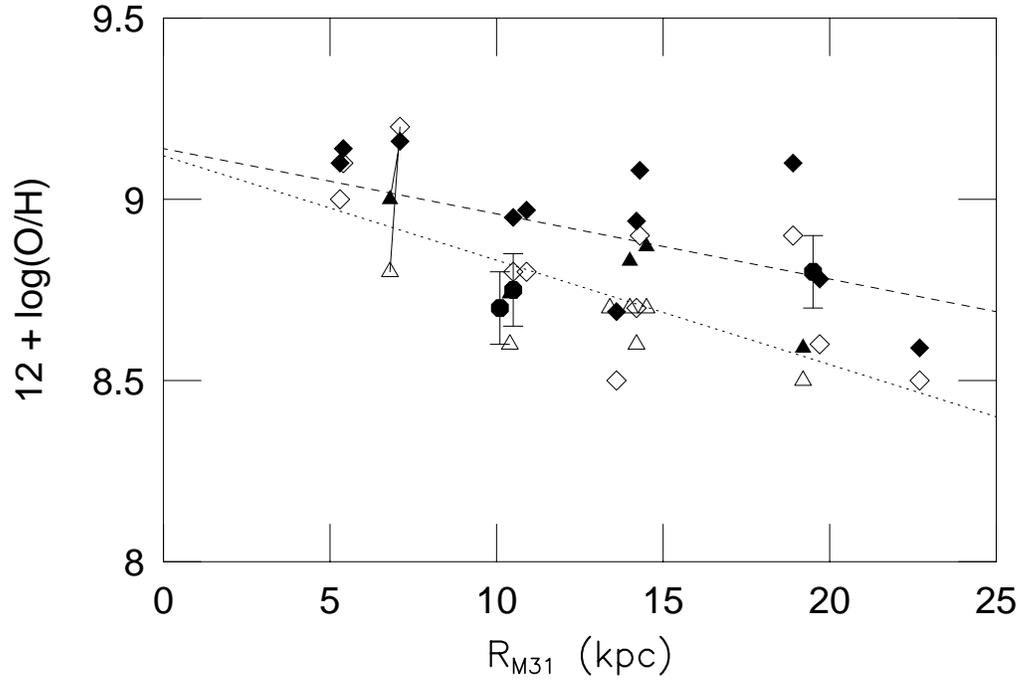}
\caption{Oxygen abundances in M31 from \ion{H}{2} regions and
the three A-F supergiants in this paper. 
{\it Hollow diamonds} are the abundances from BKC82,
{\it hollow triangles} are from Dennefeld \& Kunth (1981).
The {\it dotted line} is a least squares fit to these results.
{\it Solid diamonds} and {\it solid triangles} are recalculated
oxygen abundances from the data using the \R23\ calibration by 
ZKH94.  The {\it dashed line} is the least squares fit to the 
recalibrated data.  The results for one nebula, observed by both 
groups, is connected with a solid line (from both calibrations).  
Stellar oxygen abundances (NLTE) from this paper are plotted 
as {\it filled circles} with errorbars.
\label{m31-ograd} }
\end{figure}

Oxygen and nitrogen abundances in M31, as inferred from the
analysis of \ion{H}{2} regions have been investigated by 
Blair \etal (1982 = BKC82) and Dennefeld \& Kunth (1981).
Their results are in good agreement with each other, and in 
Fig.~\ref{m31-ograd} we plot their nebular oxygen abundances 
as a function of M31 galactocentric distance 
(\Rm31).  We also show a least squares fit to these data 
which implies a radial abundance gradient of $-$0.029~\dexkpc\ 
(intercept value 12+log(O/H) = 9.12\,dex, same as BKC82).  

Uncertainties in the slope are significant; the oxygen 
abundance drops by about a factor of 4, and yet the uncertainty
in each oxygen abundance is about a factor of 2, and the
range in the data at a given \Rm31\ is about a factor of 2 to 3.  
Thus, the oxygen radial gradient is rather 
uncertain, $-$0.029 $\pm$0.023~\dexkpc. 
Most of the uncertainties (per nebula, as well as the range
from various nebulae at the same \Rm31) come from 
uncertainties in the \R23\ empirical calibration(s) used.
BKC82 based their nebular analysis
on the calibrations by Pagel \etal (1979), although their
\ion{H}{2} regions had a larger range in metallicity and
lower excitation than those used by Pagel \etal for the
calibration.  
Vila-Costas \& Edmunds (1992) obtained a slightly higher 
gradient ($-$0.043~\dexkpc) from a recalculation of O/H from the 
published line intensities using updated \R23\ calibrations
by Edmunds \& Pagel (1984) and supplementing the high and
low metallicity regions with calibrations by
Edmunds (1989) and Skillman (1989), respectively. 
Similarily, Zaritsky, Kennicutt \& Huchra (1994 = ZKH94) 
found a more shallow gradient, $-$0.018~$\pm$0.006~\dexkpc, 
by computing mean oxygen abundances from three different 
calibrations (Edmunds \& Pagel 1984, Dopita \& Evans 1986,
and McCall \etal 1985).  Uncertainties in the oxygen abundances
are estimated as $\pm$0.2~dex from the Pagel \etal (1979) \R23\ 
calibration (Pagel \etal 1980), yet ZKH94 found that the 
dispersion between the three methods they examined dominated 
their oxygen uncertainties.

In Fig.~\ref{m31-ograd} we show the positions and (NLTE)
oxygen abundances of the three A-F supergiants analysed here.
The stellar positions 
were calculated adopting the same parameters for M31 as 
from Blair \etal (1982) for comparison purposes
(i.e., 690 kpc distance to M31, 37.5$^o$ position 
angle for the major axis, 77.5$^o$ inclination). 
The stellar abundances are consistent with the nebular 
picture, i.e., the stellar oxygen abundances lie within 
the range of nebular results, and yet there are three 
striking points to the stellar abundances.  
Firstly, the stellar abundance uncertainties are
noticably smaller than the full range in the nebular abundances;
this is probably due to uncertainties in the application of the 
\R23\ calibrations, since individual nebular abundances should
have similar uncertainties to the stellar abundances.
Secondly, the two stars near 10~kpc yield very similar 
oxygen abundances (log(O/H)$\sim$8.75), and these abundances 
are in excellent agreement with the mean nebular abundance 
from BKC82 (log(O/H)$\sim$8.7) at this \Rm31\ distance.
Thirdly, the star near 20~kpc has an abundance in excellent
agreement with the mean nebular abundance from ZKH94
(log(O/H)$\sim$8.8), but this abundance is also similar 
to those near 10~kpc.   This is interesting because
the stellar abundances suggest that there is no radial 
gradient between 10 and 20~kpc.

Stellar and nebular oxygen abundances are usually in
very good agreement.  For example, Cunha \& Lambert 
(1992) found very good agreement between the Orion 
nebular oxygen abundances and B-star abundances. 
Smartt \& Rolleston (1997) determined oxygen in B-stars
in the Galaxy and found the same radial gradient as from
nebulae.  Venn (1999) found that B-stars, A-F supergiants,
and nebulae all yield the same oxygen abundances in the
SMC.  McCarthy \etal (1995) and Monteverde \etal (1997)   
find oxygen from B-A supergiants in M33 that are in good
agreement with the oxygen gradient from nebular studies
in that galaxy.
Thus, we return to the third point above to further discuss
the nature of the oxygen gradient in M31 beyond $\sim$10~kpc.

To examine the gradient beyond 10~kpc, we need to
ascertain the reliability of the stellar and nebular abundances.
Firstly, are the stellar abundances sufficiently reliable
to examine the oxygen gradient?   We suggest that they are.
Analysis of A-F supergiants in the Galaxy and Magellanic Clouds
have found oxygen abundances in good agreement with nebular
results from similar model atmosphere analyses 
(Venn 1999, Luck \etal 1998, Hill 1997, Hill \etal 1995,
Luck \& Lambert 1992, Russell \& Bessell 1989).   
Furthermore, the \ion{O}{1}
feature analysed here in all three stars is simply not
very sensitive to the standard uncertainties in this
analysis (e.g., \teff, gravity, $\xi$, see Table~\ref{abu-unc},
also the NLTE corrections are quite consistent between
various detailed NLTE analyses and the corrections are modest). 
Secondly, how significant is the gradient reported from
the nebular abundances?   For example, we notice that the nebular
abundances show a large range in oxygen at any given \Rm31. 
This is interesting since there has been some 
discussion on the shape of abundance gradients
in spiral galaxies. 
For example, the question ``Do radial gradients flatten out?''
has been posed through observational and theoretical 
studies (e.g., Moll\'a \etal 1996, Zaritsky 1992, 
Vilchez et al 1988). 
Zaritsky (1992) predicts that the radial abundance profile
of a spiral galaxy changes slope where the rotation curve
levels off, a situation that arises from star-formation in 
a viscous disk model.
In the case of M31, one would expect this break to occur
at \Rm31 = 6 to 8~kpc (rotation curves for M31 by
Rubin \& Ford 1970, Roberts \& Whitehurst 1975).
This is not {\it clearly} seen in the nebular data, 
although neglecting the innermost \ion{H}{2} 
(with \Rm31$<$10~kpc) would result in a very flat 
gradient, possibly suggesting a two-component gradient 
as predicted.   The mean value of oxygen in the outer
disk would then be 12+log(O/H) $\sim$8.7, which is
in excellent agreement with the mean of the three 
stellar abundances presented here, log(O/H)$\sim$8.75.
 
Finally, we note that Brewer \etal (1995) also suggested
that the abundance gradient in M31 flattens out in the
outer disk.   They deduce abundances from the 
ratio of C-type (C-poor) to M-type (C-rich) AGB stars,
and included a field at \Rm31=32~kpc which had a
similar (though highly uncertain) result to their fields
near 10~kpc. 

There are many isolated A-F supergiants throughout
the disk of M31, although most tend to be near a ring
of OB associations located near \Rm31$\sim$10~kpc
(see van den Bergh 1964).  Observing more stars ($\ge$15),
especially some at \Rm31$<$10~kpc and $>$20~kpc, is desirable.
Given the small uncertainties in the oxygen abundances
from these stars, they are valuable in addressing the 
question of the nature of the oxygen abundance gradient in M31.

\subsection{Element Ratios and the Chemical Evolution of M31} 

A high resolution analysis of A-F supergiants allows us to 
determine the abundances of many elements beyond oxygen. 
Among the three stars analysed in this paper, we have also
found the abundances for several other $\alpha$-elements,
as well as iron-group and heavier elements.   

\begin{figure}
\plotone{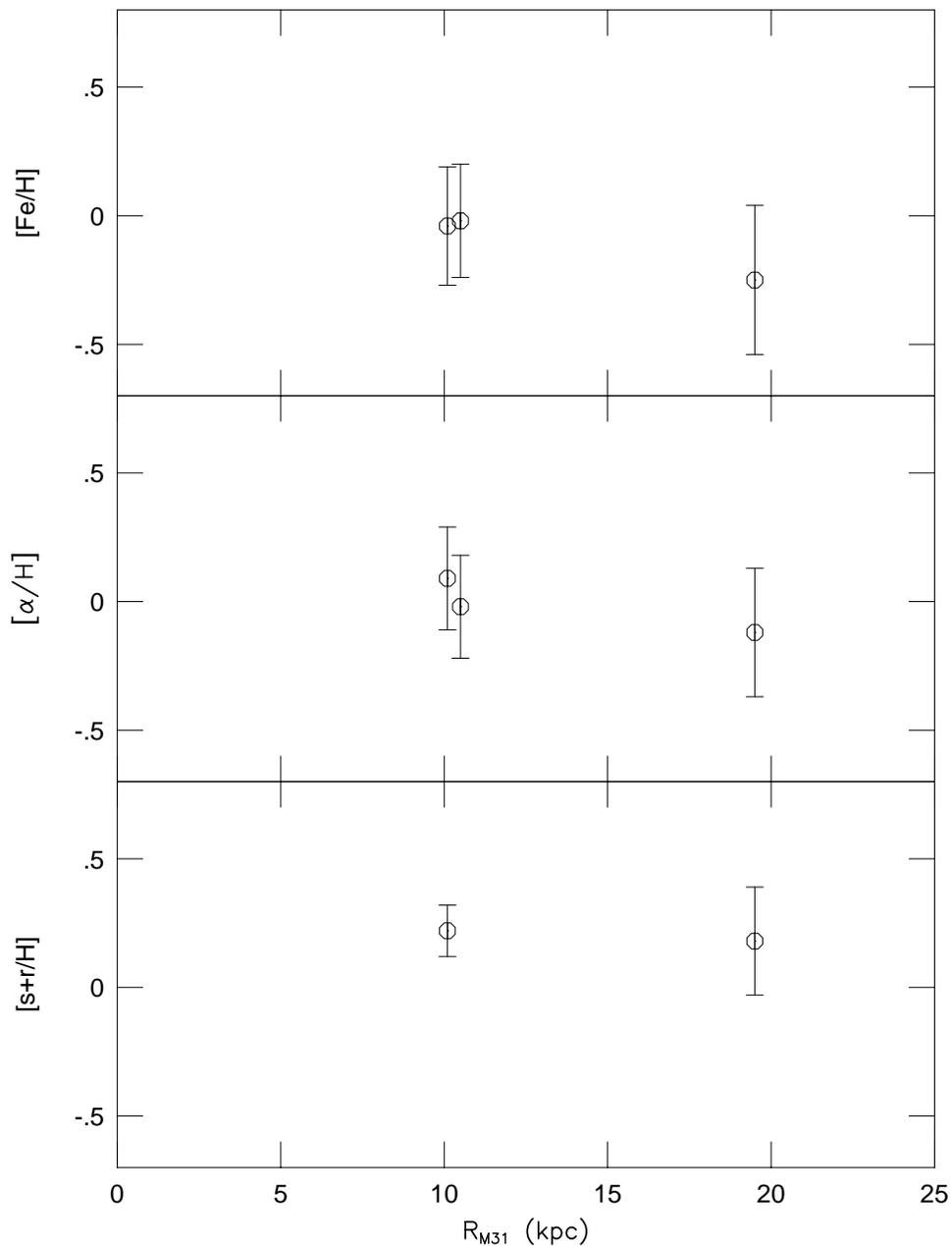}
\caption{Gradients in iron, $\alpha$-elements (except O), 
and s- and r-process elements from three A-F supergiants in M31.  
[$\alpha$/H] and [s+r/H] abundances are weighted means. 
[Fe/H] is from \feii\ for the two inner (and hotter) stars, while 
it is from \fei\ \& \feii\ for the outer star.  
We note that the [Fe/H] abundances plotted are within 0.01~dex of
the weighted mean of all of the iron-group elements per star  
(excluding \fei\ in the hotter stars since it is known to suffer from 
NLTE effects, see text). 
\label{m31-rh} }
\end{figure}

We have plotted the abundances of other elements versus 
\Rm31\ in Fig.~\ref{m31-rh}.
To reduce random errors, we have plotted a weighted mean 
of all the $\alpha$ (except O), iron-group, and 
s- and r-process elements per star. 
There is the suggestion of an abundance gradient in all
of these elements, in agreement with the {\it nebular} 
oxygen gradient.  However, the data are also consistent 
with no gradients beyond 10~kpc, as predicted in the 
viscous disk model.   It is worth noting here that 
NLTE effects have been neglected for all elements, 
other than O and Mg.   This may have an effect
on the [$\alpha$/H] and [s+r/H] plots, but should not 
affect the [Fe/H] plot.   Iron-group abundances are calculated
from the dominate ionization species in all three stars,
where NLTE effects are predicted to be quite small (see
NLTE comments above for iron).

\begin{figure}
\plotone{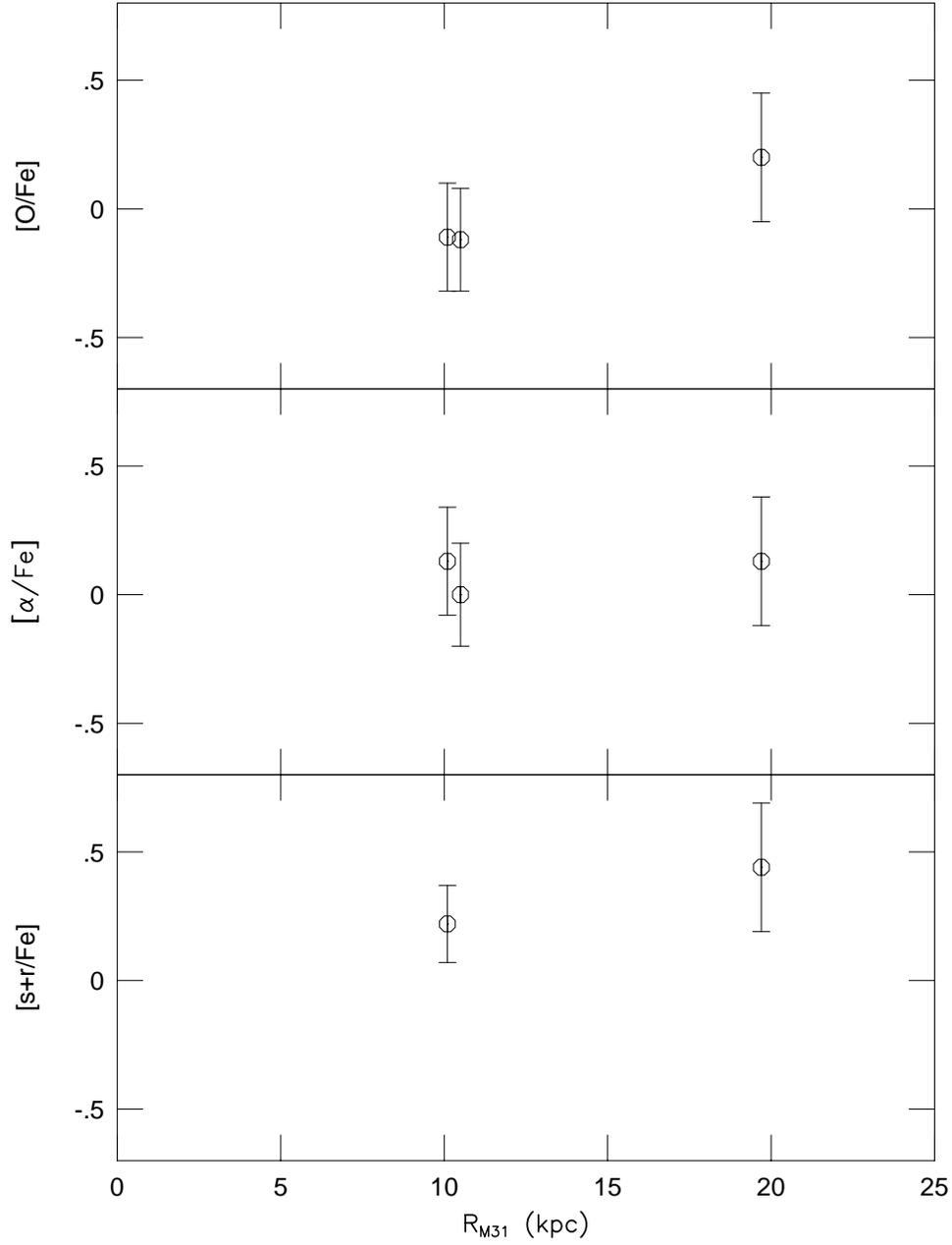}
\caption{Abundance ratios [O/Fe], [$\alpha$/Fe] (without O), 
and [s+r/Fe] from three A-F supergiants in M31.  See comments 
on weighted mean values in Fig.~\ref{m31-rh}.  Note that NLTE 
O are used. 
\label{m31-rfe} }
\end{figure}

A potentially more interesting and valuable constraint comes
from element ratios, e.g., [O/Fe], particularly versus \Rm31\
as shown in Fig.~\ref{m31-rfe}.
Element ratios are useful because they are very sensitive 
to assumptions made in chemical evolution models (e.g., 
star formation rates, the IMF, stellar yields, etc., 
see the recent review by Henry \& Worthy 1999 and the 
references cited therein).  
The proportional buildup of elements relative to one another 
depend on differences in their nucleosynthetic origins,
e.g., O/Fe depends on the star formation history because
O is produced primarily through the evolution of massive stars, 
whereas iron is ejected from all supernova events.
$\alpha$-elements have similar nucleosynthetic sources as oxygen, 
and yet the yields may vary, which affects the observed ratios, 
e.g., O/Fe versus Mg/Fe.  Meanwhile, the s- and r-process elements 
come from a very different source, intermediate-mass stars 
undergoing thermal pulsing on the AGB, and therefore are
sensitive to the IMF.  

In Fig.~\ref{m31-rfe}, [O/Fe], [s+r/Fe], and possibly 
[$\alpha$/Fe] (note that this mean $\alpha$ ratio does
not include oxygen), appear to increase in the outer disk of M31.  
We note that these increases are strongly dependent on the 
iron-group abundance in the outer most star, A-207; the 
uncertainty in the iron-group abundances for this star is 
larger than usual (as discussed above, and noted by the errorbar); 
however, we also 
believe that we have minimized potential systematic errors 
in the analysis of this star, and that the data point is 
accurate within its errorbar. 
With so few stars, the gradients (or lack of gradients) 
in these plots are not statistically significant.
However, a gradient in [O/Fe] is intriguing because, if confirmed 
by further work, then higher O/Fe in the outer disk of 
M31 could suggest, e.g., a recent burst of star formation or 
a change in the IMF.   Clearly, more stars need to be observed 
to discuss this further.

Finally, we note that the oxygen and iron abundance uncertainties
are quite similar in the early A-type supergiants when oxygen is 
determined from the \ion{O}{1} $\lambda$6156 feature and iron from 
\ion{Fe}{2} lines (see Table~\ref{abu-unc}). 
This means that systematic errors in the model atmospheres analysis
are reduced when the O/Fe {\it ratio} is examined.   This is not true 
for the cooler F-supergiants though, thus the most accurate investigation 
of [O/Fe] in M31 would concentrate on early- to mid-A supergiants only.
Unfortunately, securing enough of these targets, over a narrow range 
in temperature, and yet a large range in galactocentric radii, could
be difficult.

\subsection{Cepheid Metallicities \& Reddenings} \label{cepheids}

A-F supergiant analyses are relevant to Cepheid distance 
determinations.  These stars have similar young ages and 
intermediate-masses, and therefore should have very similar 
compositions and galactic (thin disk) locations to those of 
the Cepheids.    A-F supergiant atmospheric analyses provide
direct elemental abundances, as well as 
local reddening estimates.

\subsubsection{\it Metallicities}

The true effects of metallicity on the Cepheid PL 
relationship remain uncertain, but are expected to
affect Cepheid distances at less than the 10\% level
in M31 and Hubble Key Project galaxies 
(see discussion by Kennicutt \etal 1998). 
Metallicity is important because it increases line blanketing, 
which affects the Cepheid color calibrations and their mean 
magnitudes e.g., brighter mean magnitudes and redder colors are 
predicted at higher abundances. 

In Kennicutt \etal's (1998) discussion of the metallicity effect, 
they examined Cepheids in two fields in M101, a spiral galaxy with
a steep oxygen abundance gradient, and reviewed the results
from studies in other galaxies.
The conclusions from all of the studies suggests an  
uncertainty in the distance modulus ($\mu$) of 
$\Delta\mu\sim-0.25~\pm0.2$~\magdex\ in $VI$ luminosity.
They conclude that this relationship between $\mu$ and 
metallicity is consistent with theory
($\Delta\mu\sim-0.1$~\magdex\ in $VI$ luminosity), 
but they also note that it is consistent with no 
dependence at all.

In M31, Freedman \& Madore (1990, FM90) examined the 
Cepheid distance determinations from $BVRI$ photometry in 
Baade Fields I, III, and IV (Baade \& Swope 1963).  
They adopt galactocentric radii for these fields of 
$\sim$3, 10 and 20 kpc, with metallicities of 
0.2, 0.0 and $-$0.5~dex, respectively, from nebular 
oxygen analyses by Blair \etal (1982).  
FM90 reported $\Delta\mu=-0.32~\pm0.21$ \magdex, 
and concluded no significant metallicity dependence.
Kennicutt \etal (1998) reanalysed the FM90 data by 
recalculating the abundance gradient in M31 (see
discussion above), and found a much stronger 
metallicity dependence of $\Delta\mu=-0.94~\pm0.78$ \magdex. 
This slope would be significant if the uncertainty were
smaller; as is, the result in inconclusive.
Additionally, this neglects the question of the shape of the
abundance gradient in M31 (discussed above), which may show no 
significant range in metallicity at all beyond $\sim$10~kpc.  

With respect to our A-F supergiants, we note that the standard assumption 
is that Cepheid iron-group abundances (metallicity) will scale as the 
nebular oxygen abundances.   Clearly, this neglects potential effects 
due to the chemical evolution of the galaxy, i.e., iron and oxygen have 
different nucleosynthetic sites and thus timescales for formation and 
distribution in a galaxy.   In the LMC and SMC, this is not a problem 
since analyses of A-K supergiants have found that the O/Fe ratios are 
the same as Galactic A-K supergiants (Venn 1999, Hill 1997, 
Luck \etal 1998, Hill \etal 1995, Luck \& Lambert 1992, 
Russell \& Bessell 1989).  However, this may not
be true in M31.   The preliminary work in this paper suggests that
the Fe gradient is similar to the reported O {\it nebular} gradient, 
although both the nebular and stellar data could be consistent with 
no gradients beyond 10~kpc.
Determination of the actual iron abundances in the disk of 
M31 would allow for a proper test of metallicity effects on
the Cepheid PL relationship.

\subsubsection{\it Reddening} 

One of the more significant sources of uncertainty in Cepheid distance 
calculations today is simply the reddening estimate; in particular how 
changes in metallicity can affect Cepheid colors and cause incorrect 
reddening estimates derived from multi-wavelength Cepheid photometry. 
A-F supergiants can play a valuable role since these stars have similar 
thin-disk locations as Cepheids.   Local reddening estimates from
A-F supergiants should be similar for nearby Cepheids, assuming that
the mean reddening does not vary too widely within small regions
of M31; or, at least, they can provide a check on the Cepheid values.   
Also, local stellar values of reddening should be more accurate than 
using global reddening laws.

In this paper, we deduce the reddening to our targets using the 
photometry listed in Table~\ref{atms} and the 
spectral type-to-($B-V$)$_o$ calibration by Fitzgerald 
(1970, =F70), and the $UBV$ colors deduced from the 
adopted ATLAS9 model atmospheres 
(program UBVBUSER\footnote{Program available from R.\,L.\,Kurucz 
at {\it http://cfaku5.harvard.edu/programs.html}}). 
Uncertainties in the F70 values noted below are estimated from 
the calibration scale itself (thus, internal).  
Also, we note that the minimum amount of foreground reddening 
to M31 has been pegged
at E($B-V$)=0.08 by Burstein \& Heiles (1984), and more recently at
E($B-V$)=0.06 by Schlegel \etal (1998).
 
For 41-3712, the A3\,Ia spectral type is in good agreement
with our atmospheric parameters, and implies ($B-V$)$_o$=0.06 $\pm$0.04
from F70 and ($B-V$)$_o$=0.09 from the ATLAS9 model.
Thus, we find E($B-V$)=0.03 to 0.06 $\pm$0.04, in agreement 
with the foreground estimates.   This star is in close proximity 
to a Cepheid variable, V7184 in M31B, found by Kaluzny \etal (1998); 
their separation is 44'' on the sky, or 0.8 kpc in the disk.  
For field M31B, Kaluzny \etal\ have adopted a mean reddening 
of E($B-V$)=0.20   [this high reddening value is in agreement with 
the old Berkhuijsen \etal (1988) photometry, but our lower value 
uses the Magnier \etal (1992, 1993) CCD photometry]. 
This difference would produce an uncertainty of 
$\Delta\mu\sim+$0.46 (using R$_v$=3.1) in the $BV$ luminosity.

For 41-2368, the A8\,Ia spectral type agrees well with our atmospheric 
parameters, and implies ($B-V$)$_o$=0.14 $\pm$0.02 from F70 and 
($B-V$)$_o$=0.12 from the ATLAS9 model.  
These calibrations imply E($B-V$)=0.10 to 0.12 $\pm$0.08.  
This star is in close proximity to Cepheid \#75 from Magnier \etal (1997); 
their separation is 29'' on the sky, or 0.4 kpc in the disk.
We could not find a distance determination or reddening estimate 
for this star for a comparison. 

For A-207 in Field IV, the F5\,Ia spectral type is in good agreement 
with our atmospheric parameters, and implies ($B-V$)$_o$=+0.26 $\pm$0.11 
from F70 and ($B-V$)$_o$=+0.27 from the ATLAS9 model.  
Thus, E($B-V$)=+0.17 $\pm$0.09 [this value is slightly higher than
those found from previous studies, e.g., Baade \& Swope (1963) 
adopted E($B-V$)=0.16 $\pm$0.03, and van den Bergh (1964) reported 
E($B-V$)=0.06 $\pm$0.03 to the association OB\,184 in Field~IV].
Freedman \& Madore (1990) found E($B-V$)=0.0 to Field IV in M31
from their $BVRI$ Cepheid photometry, but this is relative to 
LMC Cepheids for which they adopted a mean reddening of E($B-V$)=0.1.  
Our higher reddening value
implies $\Delta\mu=-0.22$ in $BV$ luminosity (using R$_v$=3.1)
for their Field IV results, and a trivial correction to their
results would imply distances to Cepheids in Fields I, II, and IV 
that are in excellent agreement (24.33 $\pm$0.12, 24.41 $\pm$0.09, 
and now {\it 24.36 $\pm$0.12}, respectively).   
This result is intriguing; it may suggest no significant
metallicity dependence on Cepheid distances, assuming the
reddening to A-207 actually does reflect that to the Cepheids
in Field~IV better.  On the other hand, since our
stellar abundances indicate no significant metallicity
gradient in M31 beyond 10~kpc, then perhaps metallicity 
effects simply cannot be tested adequately in M31. 

Thus, our examination of reddening estimates in the areas 
sampled by our A-F supergiants might suggest some changes 
in Cepheid distances, ranging from $\Delta\mu\sim-$0.2 to +0.4, 
which are significant amounts.

\section{Conclusions and Future Work} 

In this paper, we have presented new abundances for 
three stars in M31, and discussed these results in the 
context of the radial gradient in oxygen observed from nebulae. 
The stellar oxygen abundances are in excellent agreement
with the nebular results, and yet they are more consistent 
with no radial gradient in oxygen (between $\sim$10 and 20~kpc).
This leads to questions on the exact form of the gradient,
e.g., does the gradient flatten out?   And/or, what is the 
dispersion in abundances at a given galactocentric distance?
We suggest that the A-F supergiants in M31 are ideal probes 
to further address these questions since the uncertainties in
their oxygen abundances are small, and there are plenty of these
stars in the disk of M31.  Observations of $\ge$15 stars, 
especially some at \Rm31$<$10~kpc and $>$20~kpc, is desirable.

We have shown that many new elemental abundances in M31 can 
be determined from stellar abundance analyses, e.g., 
present-day iron abundances.  The iron abundances presented 
here may exhibit a gradient similar to the {\it nebular} oxygen 
gradient, but we cannot discuss this in detail with observations 
of only three stars.  The constancy of a gradient between 
different elements can be valuable information for chemical 
evolution modelling of M31.  
Knowing the iron abundances in the disk of M31 could also be
valuable for Cepheid distance calibrations, particularly for
observational tests of metallicity effects on the Cepheid
PL relationship. 
 
Finally, A-F supergiant analyses are useful as reddening 
indicators, which may be valuable for accurate Cepheid 
distance determinations.   The three stars presented 
here are located near Cepheid variables in M31.  Those
Cepheids have been used to calculate distances based
on global reddening laws or direct Cepheid $BVRI$
photometry estimates.  
We find differences in the distance modulus 
of $\Delta\mu=-$0.2 to +0.4 in the $BV$ luminosity for
the few Cepheids examined here. 

Thus, detailed stellar atmosphere analyses are now possible for
individual stars in M31 and other Local Group galaxies.  
Stars serve as ideal probes of their environment, and can 
yield valuable constraints for chemical evolution models.  

\acknowledgments

KAV would like to thank Macalester College and the 
Luce Foundation for a Clare Boothe Luce professorship award.
Also, many thanks to Evan Skillman and Rob Kennicutt for 
helpful discussions and comments on the manuscript.  
JKM would like to thank the staff of the W.\,M.\,Keck
Observatory, in particular observing assistant Joel
Aycock, for efforts on the summit in support of
these HIRES observations.


\clearpage
\begin{deluxetable}{lrrrr}
\footnotesize
\tablecaption{Atmospheric Parameters for four M31 A-type Supergiants \label{atms}}
\tablewidth{0pt}
\tablehead{
\colhead{Star} & \colhead{A-207} &
\colhead{41-3712}   & \colhead{41-2368}   
& \colhead{\it 41-3654} 
} 
\startdata
$\alpha$ (J2000)\tablenotemark{a} & 00 37 45.30 & 00 45 10.36 & 00 44 16.56 & {\it 00 45 07.49} \nl
$\delta$ (J2000)\tablenotemark{a} & +39 58 23.1 & +41 36 53.6 & +41 20 59.7 & {\it +41 37 36.5} \nl   
\nl
  $V$ \tablenotemark{b}   & 17.05 $\pm$0.02 &  16.19 &    16.25  & {\it 16.47}   \nl
($B-V$) \tablenotemark{b} & +0.44 $\pm$0.03 & +0.29  &   -0.15   & {\it +0.29}   \nl
($U-B$) \tablenotemark{b} & -0.04 $\pm$0.05 & -0.89  &   -0.63   & {\it -0.70}   \nl        
($V-R$) \tablenotemark{b} & +0.50 $\pm$0.04 & +0.90  &   +0.99   & {\it +0.93}   \nl
Sp.~Ty. \tablenotemark{b} & F5 Ia         &  A3 Ia &    A8 Ia  & {\it A2 Ia} \nl
\nl
Magnier \#              & \nodata & 364858            & 274169            & {\it 369379} \nl
$V$ \tablenotemark{c}    & \nodata & 16.506 $\pm$0.021 & 16.840 $\pm$0.095 & {\it 16.335 $\pm$0.014} \nl  
($B-V$) \tablenotemark{c} & \nodata &  0.120 $\pm$0.023 &  0.242 $\pm$0.105 & {\it 0.175 $\pm$0.021} \nl
($V-R$) \tablenotemark{c} & \nodata &  0.200 $\pm$0.022 &  0.210 $\pm$0.106 & {\it 0.222 $\pm$0.015} \nl 
\nl
\teff (K)        & 6700 $\pm$300 & 8400 $\pm$200 & 8000 $\pm$200 &  {\it 9000 $\pm$500} \nl
\logg            & 0.2 $\pm$0.2  &  0.9 $\pm$0.1 & 0.75 $\pm$0.1 &  {\it 1.0 $\pm$0.3}  \nl
$\xi$(\kms)      & 8 $\pm$2      & 8 $\pm$1      & 8 $\pm$1      &  {\it 10 $\pm$5}    \nl
$v$sin$i$ (\kms) & 20 $\pm$5     &  30 $\pm$5    & 30 $\pm$5     &  {\it 30 $\pm$5}   \nl
E($B-V$)	 & +0.17 $\pm$0.09 & 0.05 $\pm$0.04  & 0.11 $\pm$0.08  & \nodata \nl
R$_{M31}$ (kpc)  & 19.5   &  10.5  &  10.1   &  {\it 9.9 } \nl
\enddata
\tablenotetext{a}{Coordinates from the Simbad database.}
\tablenotetext{b}{Colors and spectral types from Berkhuijsen \etal (1988), 
except for A207 where the colors are the averaged values from Humphreys (1979).
Berkhuijsen \etal\ report uncertainties in V of $\pm$0.2 and in colors of 
$\pm$0.3 to 0.4.}
\tablenotetext{c}{Colors from the Magnier catalogue (Magnier \etal\ 1992, 1993).
This photometry is preferred for the analyses and reddening estimates of
41-3712 and 41-2368.
}
\end{deluxetable}


\clearpage
 
\begin{deluxetable}{llllrrrrr}
\footnotesize
\tablecaption{Line Strengths and Atomic Data of Non-Iron-Group Elements \label{lines1}}
\tablewidth{0pt}
\tablehead{
\colhead{Elem} & \colhead{RMT} & \colhead{$\lambda$ (\AA)} &  
\colhead{$\chi$ (eV)}   & \colhead{log~gf} & 
\colhead{REF\tablenotemark{\dagger}}  & 
\colhead{A-207} & \colhead{41-2368} & 
\colhead{41-3712} 
} 
\startdata
 600 &  6 & 4770.00 & 7.48  & -2.33 & op  &  48 & \nodata & \nodata   \nl
 600 &  6 & 4771.72 & 7.49  & -1.76 & op  & 158 & \nodata & \nodata   \nl
 600 & 22 & 6587.75 & 8.54  & -1.02 & op  & 110 & \nodata & \nodata   \nl
 800\tablenotemark{a} & 10 & 6155.99 & 10.74 & -0.67 & op  & \nodata & \nodata & \nodata   \nl
 800\tablenotemark{a} & 10 & 6156.78 & 10.74 & -0.45 & op  & \nodata & \nodata & \nodata   \nl
 800 & 10 & 6158.19 & 10.74 & -0.31 & op  & \nodata &  85 &   85      \nl
1200 & 11 & 4702.98 &  4.35 & -0.37 & wmC & \nodata &  59 & \nodata   \nl
1200 &  2 & 5167.33 &  2.71 & -0.86 & wmB & \nodata & 174 & \nodata   \nl
1200 &  2 & 5172.70 &  2.71 & -0.38 & wmB & \nodata & 233 &  122    \nl
1200 &  2 & 5183.62 &  2.72 & -0.16 & wmB & \nodata & 275 &  150    \nl
1200 &  9 & 5528.42 &  4.34 & -0.34 & wmC & \nodata &  38 & \nodata   \nl
1201 & 10 & 4390.72 & 10.00 & -0.53 & wsmD & \nodata & 191 &   109  \nl
1201 &  9 & 4433.99 & 10.00 & -0.90 & wsmC & \nodata &  59 &  52       \nl
1400 & \nodata & 4818.10 & 4.95 & -1.57 & wmD & 38 & \nodata & \nodata   \nl
1400 & 16 & 5948.55 & 5.08  & -1.23 & wmD &  82 & \nodata & \nodata   \nl
1400\tablenotemark{a} & 29 & 6155.22  & 5.62  & -0.40 & kp & \nodata & \nodata & \nodata   \nl 
1401 &  5 & 5041.03 & 10.07 &  0.17 & wmD & \nodata & 130 &  167      \nl
1401 &  4 & 5957.56 & 10.07 & -0.35 & wmD & \nodata &  92 & \nodata   \nl
2000 &  4 & 4455.89 & 1.90  & -0.54 & wmC & 108 & \nodata & \nodata   \nl
2000 & 35 & 4878.13 & 2.71  & -0.33 & wmC & 128 & \nodata & \nodata   \nl
2000 & 21 & 5581.98 & 2.52  & -0.71 & wmD &  66 & \nodata & \nodata   \nl
2000 & 21 & 5590.13 & 2.52  & -0.71 & wmD &  27 & \nodata & \nodata   \nl
2000 & 47 & 5857.46 & 2.93  &  0.23 & wsmD & 119 & \nodata & \nodata   \nl
2000 & 18 & 6462.57 & 2.52  &  0.31 & wmD & 125 & \nodata & \nodata   \nl
2000 & 18 & 6499.65 & 2.52  & -0.59 & wmD &  54 & \nodata & \nodata   \nl
2001 & 14 & 5307.30 & 7.53  & -0.90 & wsmD &  94 & \nodata & \nodata   \nl
2001 & 20 & 5339.19 & 8.44  & -0.05 & k88 & \nodata &  68 &  51       \nl
3901 & 22 & 4786.58 & 1.03 & -1.29 & hl & 144 & \nodata & \nodata   \nl
3901 & 20 & 5087.42 & 1.08 & -0.17 & hl & \nodata &  47 & \nodata   \nl
3901 & 20 & 5119.11 & 0.99 & -1.36 & hl & 132 & \nodata & \nodata   \nl
3901 & 20 & 5205.73 & 1.03 & -0.34 & hl & \nodata &  61 & \nodata   \nl
3901 & 27 & 5480.73 & 1.72 & -0.99 & hl & 134 & \nodata & \nodata   \nl
3901 & 34 & 5728.89 & 1.84 & -1.13 & hl &  55 & \nodata & \nodata   \nl
5601 &  1 & 4554.04 & 0.00 &  0.16 & wmA & \nodata & 150 & \nodata   \nl
5601 &  1 & 4934.10 & 0.00 & -0.16 & wmB & \nodata &  95 & \nodata   \nl
5601 &  2 & 6496.90 & 0.60 & -0.38 & wmC & \nodata &  25 & \nodata   \nl
5801 & 57 & 4486.91 & 0.29 & -0.01 & mc & 103 & \nodata & \nodata   \nl
5801 &  6 & 4593.94 & 0.69 &  0.26 & mc & \nodata & 40  & \nodata   \nl
6001 & 50 & 4462.99 & 0.55 & -0.07 & wmD &  44 & \nodata & \nodata   \nl
6001 & 75 & 5130.60 & 1.30 &  0.10 & wmD &  89 & \nodata & \nodata   \nl
6001 & 75 & 5293.17 & 0.82 & -0.15 & wmD & 118 & \nodata & \nodata   \nl
\enddata
\tablenotetext{\dagger}{Reference Key:
wm = Wiese \& Martin 1980, 
op = Hibbert \etal 1991 (O), Hibbert \etal 1993 (C), 
wsm = Wiese, Smith, \& Miles 1969,
hl = Hannaford \etal 1982,
mc = Meggers \etal 1975,
kp = Kurucz \& Peytremann 1975,
k88 = Kurucz 1988.
Capitol letters denote estimated accuracy.
}
\tablenotetext{a}{These lines have been used for
spectrum syntheses so that only their atomic data 
are reported here.}
\end{deluxetable}


\clearpage
 
\begin{deluxetable}{lll|lll}
\footnotesize
\tablecaption{Elemental Abundances in four M31 A-type Supergiants  \label{abu}}
\tablewidth{0pt}
\tablehead{
\colhead{Elem} & \colhead{Sun}   & \colhead{Gal AI}   & 
\colhead {A-207} & \colhead{41-3712} & \colhead{41-2368}  
} 
\startdata
C I   & 8.52 & 8.35 $\pm$0.21  & 8.26 $\pm$0.05 (3)  & \nodata        & 
\nodata           \nl
O I\tablenotemark{a}  & 8.83 & 8.77 $\pm$0.12  & 8.8 $\pm$0.1 (S) & 
9.0 $\pm$0.1 (S) & 8.9 $\pm$0.1 (S) \nl
{\it O I}\tablenotemark{a,b} & & & 8.8 $\pm$0.1 (S) & 
8.75 $\pm$0.1 (S) & 8.7 $\pm$0.1 (S) \nl
Mg I  & 7.58 & 7.48 $\pm$0.17  & \nodata  & 7.41 $\pm$0.05 (2) & 
7.44 $\pm$0.13 (5) \nl
Mg II & 7.58 & 7.46 $\pm$0.17  & \nodata  & 7.66 $\pm$0.06 (2) & 
7.64  (1)          \nl
{\it Mg I}\tablenotemark{b} &  & & \nodata & 7.62 $\pm$0.06 (2) 
& 7.56 $\pm$0.18 (4)  \nl
{\it Mg II}\tablenotemark{b} & & & \nodata & 7.61 $\pm$0.05 (2) & 
7.59 (1)              \nl
Si I  & 7.56 & 7.48 $\pm$0.14 & 7.43 $\pm$0.30 (3) & \nodata  & 
\nodata \nl
Si II & 7.56 & 7.33 $\pm$0.17 & \nodata & 7.65 (1) & 
7.56 $\pm$0.21 (2) \nl
Ca I  & 6.35 & 6.65 $\pm$0.19 & 6.19 $\pm$0.27 (7) & \nodata  & 
\nodata \nl
Ca II & 6.35 & 6.03 $\pm$0.26 & 6.26 (1) & \nodata  & 
\nodata \nl
Sc II & 3.10 & 3.13 $\pm$0.20 & 2.69 $\pm$0.36 (2) & 3.03 $\pm$0.25 (5)  & 
3.22 $\pm$0.24 (8)  \nl
Ti I  & 4.94 & \nodata & 4.95 $\pm$0.33 (2) & \nodata & 
\nodata  \nl
Ti II & 4.94 & 4.86 $\pm$0.25 & \nodata & 4.90 $\pm$0.21 (17) & 
5.05 $\pm$0.23 (31)  \nl
Cr II & 5.69 & 5.61 $\pm$0.23 & 5.15 $\pm$0.24 (7) & 5.67 $\pm$0.22 (17) & 
5.63 $\pm$0.24 (22)  \nl
Mn II & 5.53 & 5.81 $\pm$0.20 & \nodata & 5.59 (1) & 
5.64 (1)  \nl
Fe I  & 7.50 & 7.56 $\pm$0.24 & 7.26 $\pm$0.33 (41) & 7.55 $\pm$0.30 (3)  & 
7.75 $\pm$0.25 (24)  \nl
Fe II & 7.50 & 7.40 $\pm$0.11 & 7.23 $\pm$0.23 (11) & 7.46 $\pm$0.22 (30) & 
7.48 $\pm$0.22 (29)   \nl
Ni I  & 6.25 & 6.35 $\pm$0.15 & 6.14 $\pm$0.25 (4) & \nodata & 
\nodata  \nl
Y II  & 2.23 & \nodata & 2.35 $\pm$0.12 (4) & \nodata & 
2.47 $\pm$0.18 (2)  \nl
Ba II & 2.22 & \nodata & \nodata  & \nodata & 2.41 $\pm$0.05 (3)  \nl
Ce II & 1.63 & \nodata & 1.55 (1) & \nodata & \nodata  \nl
Nd II & 1.49 & \nodata & 1.80 $\pm$0.40 (3) & \nodata & \nodata  \nl
\enddata
\tablenotetext{a}{Oxygen abundances from syntheses (S) of the 6158 \AA\ feature.}
\tablenotetext{b}{{\it O I}, {\it Mg I}, and {\it Mg II} NLTE abundances. }
\end{deluxetable}

\clearpage


\clearpage
 
\begin{deluxetable}{lrrlrlrl} 
\footnotesize
\tablecaption{ Stellar Abundance Uncertainties \label{abu-unc}}
\tablewidth{0pt}
\tablehead{
     & \colhead{A207} & \colhead{A207} & \colhead{A207} 
     &  \colhead{41-3712} & \colhead{41-3712}    
     &  \colhead{41-2368} & \colhead{41-2368} \nl
     &  \colhead{$\Delta$\teff} & \colhead{$\Delta$log\,g} & \colhead{$\Delta\xi$} 
     &  \colhead{$\Delta$\teff} & \colhead{$\Delta$log\,g}  
     &  \colhead{$\Delta$\teff} & \colhead{$\Delta$log\,g} \nl
     &  \colhead{+300\,K} & \colhead{$-$0.2} & \colhead{$-$2 km\,s$^{\rm -1}$}
     &  \colhead{+200\,K} & \colhead{$-$0.1} 
     &  \colhead{+200\,K} & \colhead{$-$0.1} 
} 
\startdata
C I   &  +0.20  &  +0.05  &  +0.04  & \nodata & \nodata & \nodata & \nodata \nl
O I   &  -0.03  &  -0.02  &  +0.08  &  +0.17 & +0.12 & +0.17 & +0.07 \nl
Mg I  & \nodata & \nodata & \nodata & +0.48 & +0.24 & +0.51 & +0.27  \nl
Mg II & \nodata & \nodata & \nodata & +0.08 & +0.08 & +0.11 & +0.13  \nl
Si I  &  +0.28  &  +0.08  &  +0.03  & \nodata & \nodata & \nodata & \nodata \nl
Si II & \nodata & \nodata & \nodata & +0.01 & +0.05 & +0.02 & +0.02  \nl
Ca I  &  +0.37  &  +0.08  &  +0.05  & \nodata & \nodata & \nodata & \nodata \nl
Ca II &  +0.12  &  +0.01  &  +0.05  & \nodata & \nodata & \nodata & \nodata \nl
Sc II &  +0.22  &  -0.02  &  +0.06  & +0.40 & +0.17 & +0.40 & +0.11  \nl
Ti I  &  +0.37  &  +0.06  &  +0.14  & \nodata & \nodata & \nodata & \nodata \nl
Ti II & \nodata & \nodata & \nodata & +0.29 & +0.12 & +0.30 & +0.08  \nl
Cr II &  +0.13  &  -0.02  &  +0.06  & +0.16 & +0.07 & +0.22 & +0.06  \nl
Mn II & \nodata & \nodata & \nodata & +0.11 & +0.06 & +0.16 & +0.05  \nl
Fe I  &  +0.23  &  +0.08  &  +0.04  & +0.44 & +0.22 & +0.50 & +0.18  \nl
Fe II &  +0.12  &  -0.02  &  +0.05  & +0.12 & +0.05 & +0.20 & +0.04  \nl
Ni I  &  +0.32  &  +0.08  &  +0.04  & \nodata & \nodata & \nodata & \nodata  \nl
Y II  &  +0.25  &  -0.02  &  +0.07  & \nodata &  \nodata & +0.53 & +0.16 \nl
Ba II & \nodata & \nodata & \nodata & \nodata &  \nodata & +0.60 & +0.19 \nl
Ce II &  +0.38  &  +0.02  &  +0.06  & \nodata &  \nodata & +0.60 & +0.18 \nl
Nd II &  +0.38  &  +0.02  &  +0.04  & \nodata & \nodata  & \nodata & \nodata \nl
\enddata
\end{deluxetable}



\clearpage
\clearpage

\clearpage
\clearpage

\clearpage
\clearpage

\clearpage
\clearpage

\clearpage
\clearpage

\clearpage
\clearpage

\clearpage
\clearpage

\clearpage
\clearpage

\clearpage
\clearpage

\clearpage
\clearpage

\clearpage
\clearpage

\end{document}